%% file: lccf_arXiv.tex
\documentclass[a4paper]{article}
\pdfoutput=1
\usepackage{fullpage}
\usepackage[x11names]{xcolor}
\colorlet{shadecolor}{black!10}
\usepackage{framed}
\usepackage{standalone}

\usepackage[utf8]{inputenc}
\usepackage{amsmath}
\usepackage{amsthm}
\usepackage{comment}
\usepackage{amssymb}
\usepackage{authblk}

\usepackage{microtype}
  \usepackage{tikz}
  \usepackage{xspace}
  \usepackage{makecell}
  \usepackage[ruled,noline,noend]{algorithm2e}
  \usepackage{tabularx}
  \usepackage{comment}
  \usepackage{thmtools}
  \usepackage{thm-restate}
  \usepackage{verbatim}
  \usepackage{enumitem}
  \usepackage[hidelinks]{hyperref}
  \usepackage[capitalise]{cleveref}

\usepackage{tikz}
\usetikzlibrary{shapes}
\usetikzlibrary{patterns,decorations.pathmorphing, decorations.pathreplacing}
\usepackage{calc}

\newcommand\twocol[2]{%
\begin{center}%
\begin{minipage}[t]{0.5\textwidth}%
{{\centering #1}}%
\end{minipage}\hfill%
\begin{minipage}[t]{0.5\textwidth}%
{{\centering #2}}%
\end{minipage}%
\end{center}}

\newtheorem{theorem}{Theorem}
\newtheorem{fact}[theorem]{Fact}

\newtheorem{claim}[theorem]{Claim}
\newtheorem{observation}[theorem]{Observation}

\newtheorem{lem}[theorem]{Lemma}
\newtheorem{lemma}[theorem]{Lemma}
\renewenvironment{lemma}{\begin{lem}}{\end{lem}}
\crefname{lem}{Lemma}{Lemmas}
\theoremstyle{definition}
\newtheorem{definition}[theorem]{Definition}
\theoremstyle{remark}
\newtheorem{example}[theorem]{Example}

\def\dd{\mathinner{.\,.}}
\newcommand{\Oh}{\mathcal{O}}

\newcommand{\F}{\mathcal{F}}
\newcommand{\N}{\mathsf{NpFrag}}

\newcommand{\R}{\mathcal{R}}

\newcommand{\Frag}{\mathsf{Frag}}
\renewcommand{\S}{\mathsf{SYNC}}

\newcommand{\integ}{\mathbb{Z}}
\newcommand{\sync}{\mathsf{sync}}
\newcommand{\Steps}{\mathsf{Steps}}

\newcommand{\LCCF}{\mathsf{LCCF}}
\newcommand{\PRECT}{\textsc{Max-Weight Intersection of Compatible Rectangles in 4D}\xspace}

\newcommand{\FFP}{\textsc{Fragment-Families-Problem}\xspace}

\newcommand{\per}{\mathsf{per}}
\newcommand{\type}{\mathsf{type}}

\newcommand{\Pairs}{\mathsf{CAND}}
\newcommand{\LeftWin}{\mathsf{LeftWin}}

\newcommand{\RightWin}{\mathsf{RightWin}}
\newcommand{\LS}{\mathsf{LeftSync}}
\newcommand{\RS}{\mathsf{RightSync}}
\newcommand{\HPerSuf}{\mathsf{HPerSuf}}
\newcommand{\HPerPref}{\mathsf{HPerPref}}
\newcommand{\Lyndons}{\mathsf{Lyn}}

\newcommand{\Int}{\mathcal{I}}
\newcommand{\Intr}{\mathcal{J}}
\newcommand{\pstart}{\mathsf{first}}
\newcommand{\pend}{\mathsf{last}}

\newcommand{\OccAll}{\mathsf{Occ}}
\newcommand{\Occ}{\mathsf{FirstPos}}
\newcommand{\Occr}{\mathsf{LastPos}}
\newcommand{\RECT}{\mathsf{RECT}}

\newcommand{\SA}{\mathsf{SA}}
\newcommand{\PA}{\mathsf{PA}}
\newcommand{\ST}{\mathsf{ST}}

 \newcommand{\defproblem}[3]{
  \vspace{2mm}
\noindent\fbox{
  \begin{minipage}{0.96\textwidth}
  \textsc{#1}\\
  {\bf{Input:}} #2  \\
  {\bf{Output:}} #3
  \end{minipage}
  }
  \vspace{2mm}
}

\begin{document}

\title{Quasi-Linear-Time Algorithm\\ for Longest Common Circular Factor}

\author[1]{Mai Alzamel}
\author[1]{Maxime Crochemore}
\author[1]{Costas S. Iliopoulos}
\author[2]{Tomasz Kociumaka}
\author[2]{Jakub Radoszewski}
\author[2]{Wojciech Rytter}
\author[2]{Juliusz Straszy\'nski}
\author[2]{Tomasz Wale\'n}
\author[2]{Wiktor Zuba}

  \affil[1]{\normalsize Department of Informatics, King’s College London, London, UK\\
    \texttt{[mai.alzamel,maxime.crochemore,costas.iliopoulos]@kcl.ac.uk}}
    
  \affil[2]{\normalsize Institute of Informatics,  University of Warsaw, Warsaw, Poland\\
    \texttt{[kociumaka,jrad,rytter,jks,walen,w.zuba]@mimuw.edu.pl}}

\date{\vspace{-.8cm}}

\maketitle

\begin{abstract}
We introduce the Longest Common Circular Factor (LCCF) problem in which, given strings $S$ and $T$ of length $n$,
we are to compute the longest factor of $S$ whose cyclic shift occurs as a factor of $T$.
It is a new similarity measure, an extension of the classic Longest Common Factor. 
We show how to solve the LCCF problem in $\Oh(n \log^5 n)$ time.
\end{abstract}

\section{Introduction}
We introduce a new variant of the Longest Common Factor (LCF) Problem, called the Longest Common Circular Factor (LCCF) Problem.
In the LCCF problem, given two strings $S$ and $T$, both of length $n$, we seek for the longest factor of $S$ whose cyclic shift occurs as a factor of $T$.
The length of the LCCF is a new string similarity measure that is 2-approximated by the length of the LCF.
We show that the exact value of LCCF can be computed efficiently.

A linear-time solution to the LCF problem is one of the best-known applications of the suffix tree~\cite{DBLP:journals/cacm/ApostolicoCFGM16}.
Just as the LCF problem was an extension of the classical pattern matching, the LCCF can be seen as an extension of the circular pattern matching problem.
The latter can still be solved in linear time using the suffix tree and admits a number of efficient solutions based on practical
approaches~\cite{DBLP:conf/bcb/AzimIRS14,DBLP:journals/cj/ChenHL14,DBLP:journals/jda/FredrikssonG09,DBLP:books/sp/17/IliopoulosPR17,DBLP:journals/cj/LinA12,DBLP:conf/icmmi/SusikGD13},
also in the approximate variant~\cite{DBLP:journals/almob/BartonIP14,DBLP:conf/lata/BartonIP15,DBLP:journals/jea/FredrikssonN04,DBLP:journals/jea/HirvolaT17},
as well as an indexing variants~\cite{DBLP:journals/mscs/AtharBBGILP17,DBLP:books/sp/17/IliopoulosPR17,DBLP:conf/walcom/IliopoulosR08},
and the problem of detecting various circular patterns~\cite{DBLP:journals/cj/LinJA15}.
The LCCF problem is further related to the notion of unbalanced translocations~\cite{DBLP:journals/corr/abs-1812-00421,DBLP:journals/tcs/ChoHK15,UnbC,UnbB,UnbA}.

One can formally state the problem in scope as follows.

\defproblem{Longest Common Circular Factor (LCCF)}
{
  Two strings $S$ and $T$ of length $n$ each.
}{
  A longest pair of factors, $F$ of $S$ and $F'$ of $T$, for which there exist strings $U$ and $V$ such that $F=UV$ and $F'=VU$;
  we denote $\LCCF(S,T)=(F,F')$.
}

Our main result is the following.

\begin{restatable}[Main Result]{theorem}{thmmain}\label{thm:main}
The Longest Common Circular Factor problem on two strings of length $n$ can be solved in $\Oh(n\log^5 n)$ time and $\Oh(n \log^2 n)$ space.
\end{restatable}

\paragraph*{Our approach.}
We apply techniques from the area of internal pattern matching (in case $U$ and $V$ are not highly periodic; \cref{sec:cube-free}) and Lyndon roots (otherwise; \cref{sec:periodic}).
The LCCF problem is reduced to
finding configurations satisfying conjunction of four conditions of type $i\in \OccAll(X)$, where $\OccAll(X)$ is the set of occurrences matching a factor $X$.

Each configuration can be decomposed into two subconfigurations (pairs of consecutive fragments), one in $S$ and one in $T$.
We guarantee that the number of subconfigurations is nearly linear so that we can compute them all for both $S$ and $T$.
Then, the task reduces to finding
two subconfigurations which agree (produce a full configuration) and constitute
an optimal solution. This is done using geometric techniques in \cref{sec:ffp}.
Each condition $i \in \OccAll(X)$ can be seen as membership of a point in a range
since $\OccAll(X)$ form an interval in the suffix array.
This gives a reduction of the LCCF problem to an
intersection problem for 4D-rectangles.
The latter task is solved efficiently using a sweep line algorithm.

\section{Preliminaries}
We consider strings over an integer alphabet $\Sigma$.
If $W$ is a string, then by $|W|$ we denote its length and by $W[1],\ldots,W[|W|]$ its characters.
By $x=W[i \dd j]$ we denote a \emph{fragment} of $W$ between the $i$th and $j$th character, inclusively.
We also denote this fragment $x$ by $W[i \dd j+1)$, and we define $\pstart(x)=i$ as well as $\pend(x)=j$.
If $\pstart(x)=1$, then $x$ is a prefix, and if $\pend(x)=|W|$, it is a suffix of $W$.
Fragments $x$ and $y$ are \emph{consecutive} if $\pend(x)+1=\pstart(y)$; we then also say that $y$ \emph{follows} $x$.

The string $W[i] \cdots W[j]$ that corresponds to the fragment $x$ is a \emph{factor} of $W$.
We say that two fragments \emph{match} if the corresponding factors are the same.
Let us note that a fragment can be represented by its endpoints in $\Oh(1)$ space; this representation can also be used to specify the corresponding factor.

By $W^R$ we denote the reversal of the string $W$.
By $\per(W)$ we denote the shortest period of $W$.
A string $W$ is called \emph{(weakly) periodic} if its shortest period $p$ satisfies $2p \le |W|$.
Fine and Wilf's periodicity lemma~\cite{fw:65} asserts that if a string $W$ has periods $p$ and $q$ such that $p+q \le |W|$, then $\gcd(p,q)$ is also a period of $W$.

We define the \emph{type} of a (non-empty) string $U$ as $\type(U)=\lfloor{\log (|U|+1)-1}\rfloor$.
We denote by $\LCCF_{a,b}(S,T)=(F,F')$ the longest common circular factor of $S$ and $T$
such that $F=UV$, $F'=VU$, $\type(U)=a$, and $\type(V)=b$.
We also say that it is the type-$(a,b)$ LCCF.
Our strategy is to compute $\LCCF_{a,b}(S,T)=(F,F')$ independently for every pair $(a,b)$ satisfying
$0\le a,b\le \min(\type(S),\type(T))$, and afterwards we report the longest alternative (over all pairs $(a,b)$)
and of the LCF of $S$ and $T$ (corresponding to $U=\varepsilon$ or $V=\varepsilon$) as the final result.

\subsection{Synchronizing Functions}
Let $W$ be a string of length $n$.
By $\Frag_t(W)$ we denote the set of fragments of $W$ of length $t$ and by $\N_t(W)$ we denote the set of fragments of $W$ of length $t$ that have a period $\le t/3$.
By $\Frag_t(x)$ and $\N_t(x)$, we denote the subsets of comprising of fragments contained within a longer fragment $x$ of $W$. 

\begin{figure}[b!]
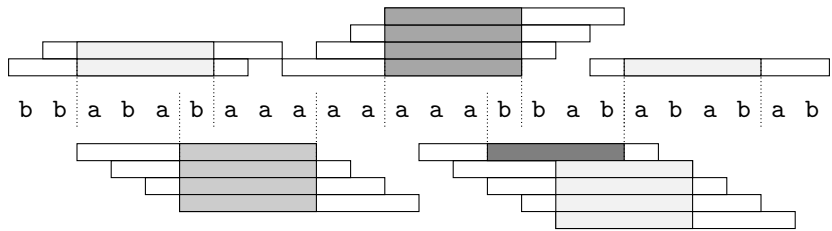

{\centering \include{rys_synch}}
\vspace*{-0.25cm}
\caption{
  For each 7-fragment, its 4-synchronizer (if not $\bot$) is shown as a shaded box.
  Observe that two 7-fragments $W[1\dd 7],\, W[15\dd 21]$ are
  matching the same factor $\mathtt{bbababa}$,
  consequently their 4-synchronizers (as fragments) start at the same relative position (the third one).
  Note that only 6 out of 21 length-4 fragments are synchronizers.
}
\label{fig:synchex}
\end{figure}

\begin{definition}[Kociumaka {\cite[Definition 4.2.1]{Kocium}}]
A function $\sync_\tau : \Frag_{2\tau-1}(W) \to \Frag_{\tau}(W) \cup \{\bot\}$ is
called \emph{$\tau$-synchronizing} (see \cref{fig:synchex}) if it satisfies the following conditions for each fragment $x \in \Frag_{2\tau-1}(W)$:
\begin{itemize}
  \item If $\sync_\tau(x)=\bot$, then $\N_\tau(x)=\emptyset$;
  \item If $\sync_\tau(x)\ne\bot$, then $\sync_\tau(x) \in \N_\tau(x)$;
  \item If two fragments $x,x'\in \Frag_{2\tau-1}(W)$ match, and
  $\sync_\tau(x) = x[s \dd s+\tau)$, then $\sync_\tau(x') = x'[s \dd s+\tau)$ for the same $s$.
  In other words, if $x=W[p \dd q]$ and $x'=W[p' \dd q']$, then $\sync_\tau(x) = W[p+s \dd p+s+\tau)$ and $\sync_\tau(x')=W[p'+s \dd p'+s+\tau)$.
\end{itemize}
\end{definition}
The elements of $\Frag_\tau(W)$ for $\tau=2^k$ are called here \emph{basic fragments}.

\begin{example}
Consider a special case of a cube-free word $W$.
Let $\mathit{DBF}(u)$ be the identifier of a $\tau$-basic fragment $u$ of $W$
in the {\it Dictionary of Basic Factors} \cite{DBLP:conf/stoc/KarpMR72} and $\pi$
be a permutation of all (linearly many) $\tau$-basic identifiers.
Each identifier is an integer in the range $[1\dd n]$. For
a fragment $x$ of size $2\tau-1$, we could define $\sync_\tau(x)$ as the first $\tau$-basic 
fragment $u$ (from the left) with minimal $\pi(\mathit{DBF}(u))$.
Then $\sync_\tau$ satisfies the conditions of a synchronizing function.
If we take a random permutation $\pi$, then it has other useful
properties in expectation. This approach can also be derandomized and generalized to arbitrary texts, as shown in the following lemma.
\end{example} 

For a function $f$ on fragments of $W$ length $t$, by $\Steps(f)$ we denote one plus the number of positions $i$ such that $f(W[i \dd i+t)) \ne f(W[i+1 \dd i+t])$.
The following fact provides an efficient construction of a $\tau$-synchronizing function with a small number of steps.
It was presented as Lemma~4.4.9\ in \cite{Kocium}; its randomized version originates from~\cite{DBLP:conf/soda/KociumakaRRW15}.

\begin{lemma}[{\cite[Lemma 4.4.9]{Kocium}}]\label{lem:Kocium}
  For a string $S$ of length $n$ and $\tau \le n/2$, in $\Oh(n)$ time one can construct a $\tau$-synchronizing function $\sync$ (stored in an array)
  such that $\Steps(\sync) = \Oh(n/\tau)$.
\end{lemma}

\noindent
The set of \emph{$\tau$-synchronizers} of a fragment $x$ is $\S_\tau(x):= \sync_\tau(\Frag_{2\tau-1}(x)) \setminus \{\bot\}$.

\section{Nonperiodic-Nonperiodic Case}\label{sec:cube-free}
We say that a string $U$ of type $a$ is \emph{highly periodic} if $\per(a)\le 2^a/3$.
We consider now $\LCCF_{a,b}(S,T)=(F,F')$
such that $F=UV$, $F'=VU$, $U$ is of type $a$, $V$ is of type $b$,
and neither $U$ nor $V$ is highly periodic.
We call it the \emph{nonperiodic-nonperiodic} case.

For a pair of fragments $(u,v)$, by
$\Gamma_{u,v}$ we denote a condition which states that $u$ is followed by a fragment that matches $v$
and by $\Delta_{u,v}$ we denote a condition which states that $v$ follows a fragment that matches $u$.
We say that two pairs of consecutive fragments, $(x,y)$ in $S$ and $(z,t)$ in $T$, \emph{agree} if and only if 
\[\Gamma_{y,z} \text{ and } \Delta_{y,z} \text{ and } \Gamma_{t,x} \text{ and } \Delta_{t,x}.\]
We reduce the LCCF problem in this case to the following abstract problem; see~\cref{fig:concrete}.

\defproblem{\FFP}{
  Two collections $\F_1$ and $\F_2$ of pairs of consecutive fragments of string $W$ of length $n$, with $m=|\F_1|+|\F_2|$
}{
  $(x,y) \in \F_1$ and $(z,t) \in \F_2$ that agree and maximize $|x|+|y|+|z|+|t|$
}

\begin{figure}[b!]
\vspace*{-.25cm}
{\centering
\include{_fig_concrete_ex}
}
\vspace*{-.75cm}
\caption{
  Pairs $(x,y)$ and $(z,t)$ agree; $txyz$ and $yztx$ form a common circular factor of $S$ and $T$.
}
\label{fig:concrete}
\end{figure}

Let us define basic fragments called the \emph{left $k$-window} and the \emph{right $k$-window}:
\begin{align*}
  \LeftWin_k(W,i)&=W[\max(1,i-2^{k+2}+2)\dd i-1],\\
  \RightWin_k(W,i)&=W[i\dd \min(|W|,i+2^{k+2}-3]).
\end{align*}

\begin{figure}[htb]
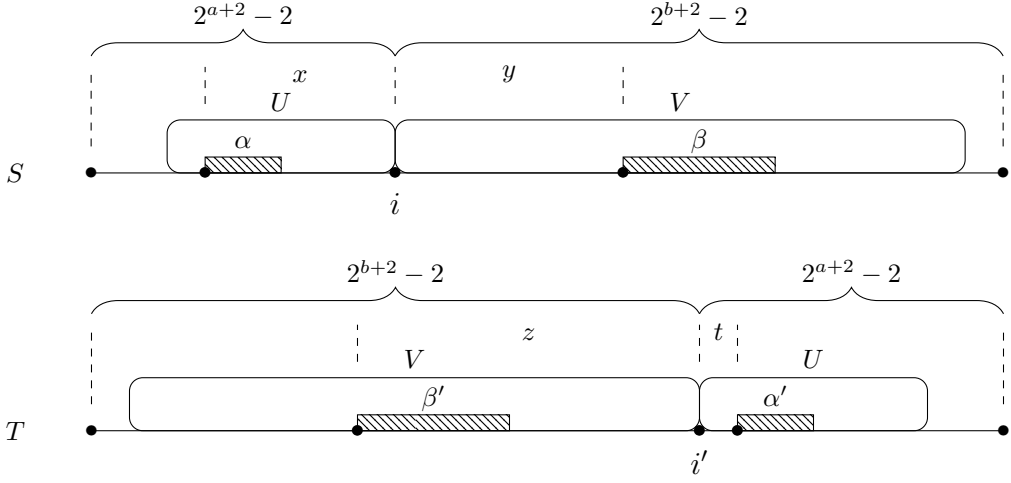

{\centering
\vspace*{-0.25cm}
\include{_fig_nonperiod_image}
\vspace*{-0.5cm}
\include{_fig_nonperiod_image2}
}
\vspace*{-.5cm}
\caption{
  Assume that $\alpha \in \LS_a(S,i),\;\beta \in \RS'_b(S,i)$, $\beta' \in \LS_a(T,i'),\;\beta' \in \RS'_b(T,i')$,
  fragments $\alpha$ and $\alpha'$ match, and fragments $\beta$ and $\beta'$ match.
  Then $\Psi_S(\alpha,i,\beta)=(x,y)$ agrees with $\Psi_T(\beta',i',\alpha')=(z,t)$ iff 
  there is a common circular factor of $S$ and $T$: $F=UV$, $F'=VU$, where $U=tx$ and $V=yz$.
}
\label{fig:crucial}
\end{figure}

\smallskip\noindent
For a string $W$ we introduce the following set of $2^{k-1}$-synchronizers:
\begin{align*}
  \LS_k(W,i)&=\S_{2^{k}}(\LeftWin_{k}(W,i)),\\
  \RS_k(W,i)&=\S_{2^{k}}(\RightWin_{k}(W,i)).
\end{align*}

By $\RS'_k(W,i)$ we denote the singleton of the leftmost fragment in $\RS_k(W,i)$ or an empty set if there is no such fragment.
For fragments $\alpha,\beta$ of $W$ and position $i$, by
\[\Psi_W(\alpha,i,\beta)= (W[\pstart(\alpha) \dd i),W[i \dd \pstart(\beta)))\]
we denote a pair of consecutive fragments of $W$ that are delimited by the starting positions of $\alpha$ and $\beta$ and the index $i$.
We then define the set of ``candidates'' (see Fig.~\ref{fig:crucial}):
$$\Pairs_{a,b}(W) = \{\Psi_W(\alpha,i,\beta)\;:\; \alpha \in \LS_a(W,i),\;\beta \in \RS'_b(W,i),\; i \in [1 \dd |W|]\}.$$

\noindent Using the terminology 
an informal scheme of a general algorithm  is as follows:

\smallskip
\begin{center}
\begin{minipage}{10cm}
\begin{shaded*}

\noindent {\bf Algorithm} {\it Compute}-$\LCCF_{a,b}(S,T)$
\begin{enumerate}

\item Compute the sets $\Pairs_{a,b}(S),\; \Pairs_{b,a}(T)$

\item Find two pairs $(x,y)\in \Pairs_{a,b}(S), (z,t)\in \Pairs_{b,a}(T)$

which agree and have maximum $|t|+|x|+|y|+|z|$

\item {\bf return} $txyz$
\end{enumerate}
\end{shaded*}
\end{minipage}
\end{center}

\begin{lemma}[Correctness for Nonperiodic-Nonperiodic Case]\label{lem:cor nn}
  The $\LCCF_{a,b}$ problem in the nonperiodic-nonperiodic case can be reduced to \FFP$(\F_S,\F_T)$
  for $\F_S=\Pairs_{a,b}(S)$ and $\F_T=\Pairs_{b,a}(T)$.
\end{lemma}
\begin{proof}
  Take a pair of fragments $f_S$ of $S$ and $f_T$ of $T$ such that $f_S$ is an occurrence of a factor $F=UV$ and $f_T$ is an occurrence of a factor $F'=VU$ 
  such that $U$ is of type $a$ and $V$ is of type $b$ and none of them is highly periodic.
  Denote by $u_S$ and $v_S$ the consecutive fragments of $f_S$ corresponding to $U$ and $V$, and similarly
  by $v_T$ and $u_T$ the consecutive fragments of $f_T$ corresponding to $V$ and $U$, and let $i=\pstart(v_S)$ and $j=\pstart(u_T)$.
  Let $u'_S$ and $u'_T$ be the leftmost fragments of $u_S$ and $u_T$ of length $2^{a+1}-1$ that are not highly periodic.
  Take a $2^{a}$-synchronizer $\sync_{2^{a}}(u'_S)$ (it belongs to $\LS_a(S,i)$ and starts at position $k$ of $u_S$),
  and $2^{a}$-synchronizer $\sync_{2^{a}}(u'_T)$ (it belongs to $\RS_a'(T,j)$, and, by the synchronization property, starts at position $k$ of $u_T$).
  Symmetrically, let $v'_S$ and $v'_T$ be the leftmost fragments of $v_S$ and $v_T$ of length $2^{b+1}-1$ that are not highly-periodic.
  The $2^{b}$-synchronizers $\sync_{2^{b}}(v'_S)$ and $\sync_{2^{b}}(v'_T)$ belong to $\RS'_b(S,i)$ and $\LS_b(T,j)$, respectively,
  and start at the same position $l$ of $v_S$ and $v_T$.
  This means that there exists a pair $(x,y)\in\Pairs_{a,b}(S)$, such that $x=u_S[k\dd |U|]$ and $y=v_S[1\dd l)$, and a pair $(z,t)\in\Pairs_{b,a}(T)$, such that
  $z=v_T[l\dd |V|]$ and $t=u_T[1\dd k)$, which agree as $y$ is followed by $v_S[l\dd |V|]$, which matches $z$, $z$ is preceded by $v_T[1\dd l)$ which matches $y$,
  $t$ is followed by $u_T[k\dd |U|]$ which matches $x$ and $x$ is preceded by $u_S[1\dd k)$ which matches $t$.

  Conversely for two pairs $(x,y)\in\Pairs_{a,b}(S),(z,t)\in\Pairs_{b,a}(T)$ that agree there exists a factor $F$ in string $S$ matching $txyz$
  and a factor $F'$ matching $yztx$ in $T$.
  Thus, there is a one-to-one correspondence between pairs that agree and fragments of strings of right type that are cyclic shifts.
  Hence by finding two pairs that agree and maximize $|x|+|y|+|z|+|t|$ we find a solution to $\LCCF_{a,b}$ problem.
\end{proof}

\begin{lemma}[Complexity for Nonperiodic-Nonperiodic Case]\label{lem:nn}
  In the nonperiodic-nonperiodic case the LCCF problem can be reduced in $\Oh(n\log^2 n)$ time to $\Oh(\log^2 n)$ instances of \FFP\ with $m=\Oh(n)$.
\end{lemma}
\begin{proof}
  We use the reduction of \cref{lem:cor nn}.
  \begin{claim}
    For a string $W$ of length $n$ and integer $k \le \type(W)$,
    \[\sum_i |\LS_k(W,i)|=\Oh(n),\qquad\sum_i |\RS_k(W,i)| = \Oh(n).\]
    Consequently, for any integers $a,b \le \type(W)$, $|\Pairs_{a,b}(W)|=\Oh(n)$.
  \end{claim}
  \begin{proof}
    For a given $\tau=2^{k}$, each $\tau$-fragment $\alpha$ can belong to only $\Oh(\tau)$ sets $\LS_k(W,i)$.
    Moreover, $\bigcup_i \LS_k(W,i) = \Oh(n/\tau)$ by \cref{lem:Kocium}.
    This yields the first part of the claim.

    Finally, $|\Pairs_{a,b}(W)| \le \sum_i |\LS_a(W,i)|$.
  \end{proof}

  By the claim, $m = \Oh(n)$ in every instance of \FFP.

  For each $\tau=2^k$, we compute a $\tau$-synchronizing function $\sync_\tau$ using \cref{lem:Kocium}.
  This takes $\Oh(n \log n)$ time in total.
  The sets $\LS_k(W,i)$ and $\RS_k(W,i)$ (and, thus, $\RS'_k(W,i)$) can be computed for any $k \le \type(W)$ in linear time
  using a sliding window approach.
  From them we can compute any set $\Pairs_{a,b}(W)$ straight from definition.
  By the claim, over all $a,b \le \type(W)$ the complexity is $\Oh(n \log^2 n)$.
\end{proof}

\section{Periodic-Periodic Case}\label{sec:periodic}
We consider now $\LCCF_{a,b}(S,T)=(F,F')$
such that $F=UV$, $F'=VU$, $U$ is of type $a$, $V$ is of type $b$,
and both $U$ and $V$ are highly periodic.

Recall that a \emph{Lyndon string} is a string that is lexicographically smaller than all its proper suffixes.
If $W$ is a weakly periodic string with the shortest period $p$, then its \emph{Lyndon root} $\lambda$ is the Lyndon string that is a cyclic shift of $W[1 \dd p]$.
A Lyndon representation of $W$ is then $(c,e,d)$ such that $W=\lambda'\lambda^e\lambda''$ where $|\lambda'|=c<|\lambda|$, $|\lambda''|=d<|\lambda|$; see~\cite{DBLP:journals/tcs/CrochemoreIKRRW14}.
Lyndon strings have the following synchronization property that follows from the periodicity lemma:
if $\lambda$ is a Lyndon string, then it has exactly two occurrences in $\lambda^2$; see \cite{AlgorithmsOnStrings}.

\begin{figure}[h]
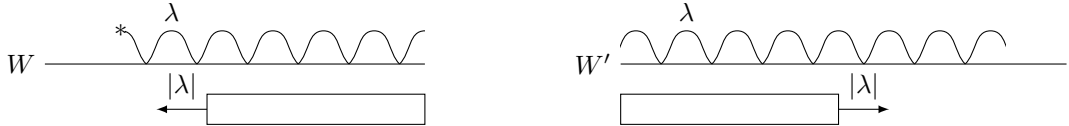

\vspace*{-0.1cm}
\twocol{\include{_rys1}}{\include{_rys2}}
\vspace*{-0.6cm}
\caption{
  Illustration of \cref{obs:lambda}.
  A highly periodic suffix of $W$ that is also a prefix of $W'$ of length at most $\min(|X|,|Y|)-|\lambda|$ can be extended by $|\lambda|$ characters.
}
\label{fig:JKS1}
\end{figure}

For a string $W$, by $\HPerPref(W)$ and $\HPerSuf(W)$ we denote the longest highly periodic prefix and suffix of $W$.
Let us start with the following simple observation; see Fig.~\ref{fig:JKS1}.

\begin{observation}\label{obs:lambda}
  Let $W$ and $W'$ be two strings for which the strings $X=\HPerSuf(W)$ and $Y=\HPerPref(W')$ have the same Lyndon root $\lambda$.
  Then the longest suffix of $W$ that is also a prefix of $W'$ has length greater than $\min(|X|,|Y|)-|\lambda|$.
\end{observation}

\begin{figure}[t]
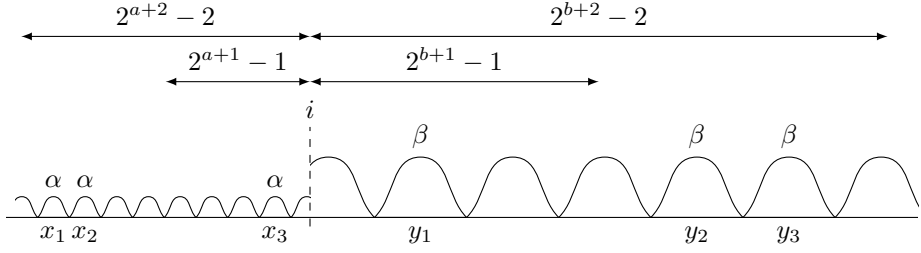

\vspace*{-0.5cm}
{\centering
\include{_rys11}
}
\vspace*{-1.25cm}
\caption{
  In this case, $\Pairs_{a,b}(W)$ contains $\Psi_W(x_p,i,y_q)$ for $p,q\in\{1,2,3\}$.
}
\label{fig:JKS2}
\end{figure}

For a fragment $u$ denote by
$\Lyndons(u)$ the set of fragments in $u$ 
corresponding to the first/second/last occurrence of Lyndon root in $\HPerSuf(u)$
and by $\Lyndons'(u)$ the set of fragments in $u$ 
corresponding to the first/second to last/last occurrence of Lyndon root in $\HPerPref(u)$.
We can redefine (see Fig.~\ref{fig:JKS2})
\begin{multline*}
  \Pairs_{a,b}(W)= \{ \Psi_W(x,i,y)\; :\; x \in \Lyndons(\LeftWin_a(W,i)),\, y \in \Lyndons'(\RightWin_b(W,i)),\,i \in [1 \dd |W|]\}.
\end{multline*}

\begin{lemma}[Correctness for Periodic-Periodic Case]\label{lem:ppcorr}
  The $\LCCF_{a,b}$ problem in the periodic-periodic case can be reduced to \FFP$(\F_S,\F_T)$
  for the redefined sets $\F_S=\Pairs_{a,b}(S)$ and $\F_T=\Pairs_{b,a}(T)$.
\end{lemma}

\begin{proof}
  Take a pair of fragments $f_S$ of $S$ and $f_T$ of $T$ such that $f_S$ is an occurrence of a factor $F=UV$ and $f_T$ is an occurrence of a factor $F'=VU$
  ($(F,F')=\LCCF_{a,b}(S,T)$)
  such that $U$ is of type $a$, $V$ is of type $b$, and both $U$ and $V$ are highly periodic.
  Denote by $u_S$ and $v_S$ the consecutive fragments of $f_S$ corresponding to $U$ and $V$, and similarly
  by $v_T$ and $u_T$ the consecutive fragments of $f_T$ corresponding to $V$ and $U$, and let $i=\pstart(v_S)$ and $j=\pstart(u_T)$.

  Let $X=\HPerSuf(\LeftWin_a(S,i))$ and $Y=\HPerPref(\RightWin_a(T,j))$.
  Note that $u_S$ is a highly periodic suffix of $\LeftWin_a(S,i)$ and that $X$ has the same period as $U$
  (a different period would contradict the periodicity lemma).
  Symmetrically, $u_T$ is a highly periodic prefix of $\RightWin_a(T,j)$ and $Y$ has the same period as $U$.
  Let $\lambda$ be the Lyndon root of $U$, and observe that $\lambda$ is also the Lyndon root of $X$ and $Y$.
  By \cref{obs:lambda}, we have
  \[|X|-|U|<|\lambda| \text{ or } |Y|-|U|<|\lambda|.\]
  as otherwise we would be able to find a Common Circular Factor
  of type $(a,b)$ that is longer by $|\lambda|$, thus contradicting our choice of $f_S$ ad $f_T$.

  If $|X|-|U|<|\lambda|$, then the first occurrence of $\lambda$ in $u_S$
  is also the first or second occurrence of Lyndon root in $X$.
  This is due to the synchronization property of Lyndon strings.
  Moreover, the first $\lambda$ in $u_T$ is also the first occurrence of $\lambda$ in $Y$.
  On the other hand, if $|Y|-|U|<|\lambda|$, then the last occurrence of $\lambda$ in $u_T$
  is the last or the second to last occurrence of $\lambda$ in $u_T$, whereas the last occurrence of $\lambda$ in $u_S$ is also the last occurrence
  of $\lambda$ in $X$. In either case, $u_S$ and $u_T$ contain a pair of corresponding occurrences of $\lambda$,
  which belong to $\Lyndons(\LeftWin_a(S,i))$ and $\Lyndons'(\RightWin_a(T,j))$, respectively; see Fig.~\ref{fig:JKS3}.

  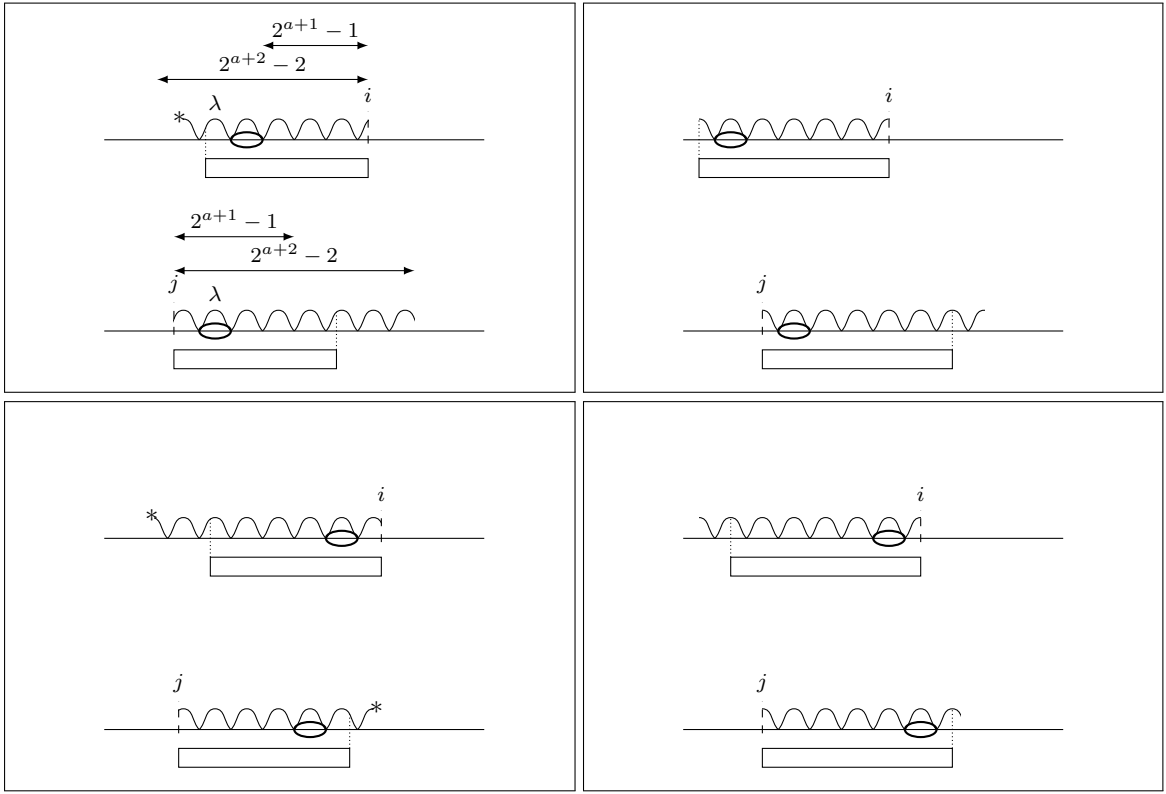
\begin{figure}[t]
    {\centering
      \label{ fig7} 
      \fbox{
        \begin{minipage}[b]{0.45\linewidth}%
            \centering\vspace{-3mm}%
            \input{rys3}
            \input{rys4}\vspace*{2mm}
        \end{minipage}
      }\;\vspace*{.75mm}%
      \fbox{
      \begin{minipage}[b]{0.45\linewidth}%
        \centering\vspace{-3mm}%
        \input{rys5}
        \input{rys6}\vspace*{2mm}
      \end{minipage}
      } 
    \fbox{
      \begin{minipage}[b]{0.45\linewidth}
        \centering\vspace{-3mm}%
        \input{rys7}
        \input{rys8}\vspace*{2mm}
      \end{minipage}
      }\;%
      \fbox{
      \begin{minipage}[b]{0.45\linewidth}
        \centering\vspace{-3mm}%
        \input{rys9}
        \input{rys10}\vspace*{2mm}
      \end{minipage}
      }
    }
    \caption{
      Four cases from \cref{lem:ppcorr}.
    }
    \label{fig:JKS3}
    \end{figure}

  As the same reasoning can be applied to $v_S$ and $v_T$, there exists a pair $(x,y)\in\Pairs_{a,b}(S)$ and a pair $(z,t)\in\Pairs_{b,a}(T)$, which corresponds
  to our choice of occurrences of Lyndon roots.
  These pairs agree and $|x|+|y|+|z|+|t|=|F|$, thus \FFP$(\F_S,\F_T)$ will find a solution at least that good.
  
  The converse direction is identical to the one from the proof of Lemma~\ref{lem:cor nn}.
\end{proof}

We proceed with an efficient implementation.
A \emph{run} in string $W$ is a maximal weakly periodic fragment $W[i \dd j]$ with a given period $p$.
We use \emph{2-period queries} which, given a weakly periodic fragment $u$ of a string, compute its shortest period and the run of the same period it belongs to.
Such queries can be answered in $\Oh(1)$ time after $\Oh(n)$-time preprocessing~\cite{DBLP:conf/soda/KociumakaRRW15}
(for a simplified solution, see~\cite{DBLP:journals/siamcomp/BannaiIINTT17}).
Let us also recall that the Lyndon representation of a run can be computed in constant time
after linear-time preprocessing~\cite{DBLP:journals/tcs/CrochemoreIKRRW14}.

\begin{lemma}[Complexity for Periodic-Periodic Case]\label{lem:pp}
  In the periodic-periodic case the LCCF problem can be reduced in $\Oh(n\log^2 n)$ time to $\Oh(\log^2 n)$ instances of \FFP\ with $m=\Oh(n)$.
\end{lemma}
\begin{proof}
For any integers $a,b \le \type(W)$, we can in $\Oh(n)$ time answer which of the fragments $\LeftWin_a(S,i)$, $\RightWin_b(S,i)$, $\LeftWin_b(T,j)$, $\RightWin_a(T,j)$ contain
highly periodic prefixes/suffixes and find the Lyndon representation of the longest such prefixes/suffixes using the following claim.

\begin{claim}
  After $\Oh(n)$-time preprocessing, $\HPerPref(u)$ and $\HPerSuf(u)$ for a type-$a$ fragment $u$ of $S$ or $T$ as well as its Lyndon representation can be computed in $\Oh(1)$ time.
\end{claim}
\begin{proof}
  To compute $\HPerPref(u)$, it suffices to ask a 2-period query for $u[1 \dd 2^{a+1}-1]$ (see~\cite{DBLP:conf/soda/KociumakaRRW15,DBLP:journals/tcs/CrochemoreIKRRW14}),
  compute the Lyndon representation of the resulting run (see~\cite{DBLP:journals/tcs/CrochemoreIKRRW14}), and then trim its Lyndon representation to $u$.
  A symmetric solution works for $\HPerSuf(u)$.
\end{proof}

\noindent
This allows us to compute the sets $\Lyndons$ and $\Lyndons'$.
Finally, $|\Pairs_{a,b}(W)|=\Oh(n)$.
\end{proof}

\section{Nonperiodic-Periodic Case}\label{sec:both}
Finally we consider the case that $\LCCF_{a,b}(S,T)=(F,F')$
such that $F=UV$, $F'=VU$, $U$ is of type $a$, $V$ is of type $b$,
and either $U$ or $V$ is highly periodic.
This case can be reduced to \FFP\ directly by combining the techniques of the previous two sections.

\begin{lemma}[Correctness for Nonperiodic-Periodic Case]
  The $\LCCF_{a,b}$ problem in the nonperiodic-periodic case can be reduced to \FFP$(\F_S,\F_T)$.
\end{lemma}
\begin{proof}
  In the proofs of Lemmas \ref{lem:cor nn} and \ref{lem:ppcorr} the $U$ and $V$ parts of the factors were considered separately.
  Hence it is enough to define $\Pairs_{a,b}(W)$ as 
  \begin{multline*}
  \{\Psi_W(\alpha,i,x)\;:\; \alpha \in \LS_a(W,i),\, x \in \Lyndons'(\RightWin_b(i)),\,i \in [1 \dd |W|]\}\: \cup\\
  \{\Psi_W(x,i,\alpha)\; :\; x \in \Lyndons(\LeftWin_a(i)), \;\alpha \in \RS'_b(W,i),\; i \in [1 \dd |W|]\}
  \end{multline*}
  and depending on which fragment is highly periodic consider one part for $\Pairs_{a,b}(S)$ and the other for $\Pairs_{b,a}(T)$. 
\end{proof}

\begin{lemma}[Complexity for Nonperiodic-Periodic Case]\label{lem:np}
  In the nonperiodic-periodic case the LCCF problem can be reduced in $\Oh(n\log^2 n)$ time to $\Oh(\log^2 n)$ instances of \FFP\ with $m=\Oh(n)$.
\end{lemma}
\begin{proof}
  The families can be computed combining the methods from Lemmas \ref{lem:nn} and \ref{lem:pp}, obtaining the desired complexities and sizes.
\end{proof}

\section{Solution to \FFP}\label{sec:ffp}
In this section we show how to solve the \FFP\ for a string $W$ of length $n$ by a reduction to intersecting
special 4-dimensional rectangles.

\noindent
First we give a geometric interpretation of two predicates:
\begin{itemize}
\item factor $U$ has an occurrence in $W$ starting at position $q$ (is a prefix of the suffix starting at position $q$)
\item and $U$ has an occurrence ending at position $q$ (is a suffix of a prefix ending at position $q$)
\end{itemize}
to the membership of $q$ in a corresponding subinterval of $[1 \dd n]$.

Let us recall that the suffix array of string $W,$ $\SA_W$, is a permutation of $[1 \dd n]$ such that
$W[\SA_W[i]\dd n]$ $< W[\SA_W[i+1] \dd n]$ for every $i \in [1 \dd n-1]$.
By $\Occ(U)$ let us denote the set of starting positions of occurrences of $U$ in $W$.
Our geometric interpretation is possible due to the following well known fact (see~\cite{AlgorithmsOnStrings}).

\begin{observation}\label{obs:SAint}
  $\Occ(U)$ is a set of consecutive elements in $\SA_W$.
\end{observation}

\smallskip
Let $\Occr(U)$ be the set of ending positions of occurrences of $U$ in $W$.
We also use the $\Occ$, $\Occr$ notation for fragments which means operations on corresponding factors.

\begin{observation}\label{obs:GDOcc}\mbox{ \ }
  \vspace*{-0.1cm}
\begin{enumerate}
\item A fragment $u$ is a prefix/suffix of the suffix starting
(prefix ending) at position $q$ iff
$q\in \Occ(u)$, $q\in \Occr(u)$ 
respectively.
\item
$\Gamma_{u,v} \equiv ((\pend(u)+1) \in \Occ(v))$ and
$\Delta_{u,v} \equiv ((\pstart(v)-1) \in \Occr(u))$.
\end{enumerate}
\end{observation}

We define a \emph{$d$-rectangle} ($d \ge 2$) as a Cartesian product of 
$d$ closed intervals, such that at least $d-2$ of them are singletons.
E.g., $\{3\} \times [2\dd 5] \times [1\dd 7] \times \{0\}$ is a 4-rectangle. 
In other words, a $d$-rectangle is an isothetic hyperrectangle of 
dimension at most 2.

By $\Int(U)$ and $\Intr(U)$ we denote the subintervals of $[1 \dd n]$ that correspond to the
intervals of $\Occ(U)$ in the suffix array $\SA_W$ and of $\Occr(U)$ in the (analogously defined) prefix array of $W$, $\PA_W$, respectively, as stated in \cref{obs:SAint}.
($\PA_W$ is a permutation of $[1 \dd n]$ such that $W[1 \dd \PA_W[i]] < W[1 \dd \PA_W[i+1]]$ for every $i \in [1 \dd n-1]$.)
For pairs $(x,y)$ and $(z,t)$ of consecutive fragments we denote:
\begin{align*}
  \RECT(x,y)\;&=\; \,\Int(x)\times \Intr(y) \times \{\SA^{-1}_W[\pend(y)+1]\}\times \{\PA^{-1}_W[\pstart(x)-1]\},\\
  \RECT'(z,t)\;&=\; \,\{\SA^{-1}_W[\pend(t)+1]\}\times \{\PA^{-1}_W[\pstart(z)-1]\}\times \Int(z)\times \Intr(t).
\end{align*}
Now \cref{obs:GDOcc}.2 implies the following.

\begin{observation}\label{obs:red}
Two pairs of consecutive fragments $(x,y)$, $(z,t)$ agree iff $\RECT(x,y)\cap \RECT'(z,t)\ne \emptyset$.
\end{observation}

Two $d$-rectangles $[a_1\dd b_1] \times \dots \times [a_d\dd b_d]$ and 
$[a'_1\dd b'_1] \times \dots \times [a'_d\dd b'_d]$
are called \emph{compatible} if, for each $i \in \{1,\ldots,d\}$, 
$[a_i\dd b_i]$ or $[a'_i\dd b'_i]$ is a singleton.
Let us note that the 4-rectangles in the above observation are compatible.

\subsection{Intersecting 4D Rectangles}
We consider two families of 4-rectangles with weights and wish to find a pair of intersecting rectangles, one per family, with maximum total weight.
The general problem of finding such an intersection of two families of $m$ weighted hyperrectangles in $d$ dimensions can be solved in $\Oh(m \log^{2d}m)$ time
by an adaptation of a classic approach~\cite{Edel}.
Below we consider a special variant of the problem that has a much more efficient solution.

\defproblem{\PRECT}
{
  Two families $\R_1$ and $\R_2$ of 4-rectangles in $\integ^4$ with integer weights containing $m$ rectangles in total,
  such that each $R_1 \in \R_1$ and $R_2 \in \R_2$ are compatible
}{
  Check if there is an intersecting pair of 4-rectangles $R_1 \in \R_1$ and $R_2 \in \R_2$ and, if so,
  compute the maximum total weight of such a pair
}

A very similar problem was considered as Problem~3\ in \cite{DBLP:journals/fuin/GrabowskiKR18} for an arbitrary $d$.
The sole difference is that the weight of an intersection of two $d$-rectangles $R_1 \in \R_1$ and $R_2 \in \R_2$
in that problem was the maximum $\ell_1$-norm of a point in $R_1 \cap R_2$.
A solution to Problem~3 for $d=4$ in the case that the 4-rectangles are compatible working in $\Oh(m \log^3 m)$ time and $\Oh(m \log^2 m)$ space
was given as \cite[Lemma 5.8]{DBLP:journals/fuin/GrabowskiKR18}.
The algorithm presented in that lemma actually solves the \PRECT\ problem and applies it for specific weight assignment of the 4-rectangles on the input.
It uses hyperplane sweep and a variant of an interval stabbing problem.
Henceforth we will use the following result.

\begin{fact}[{\cite[see Lemma 5.8]{DBLP:journals/fuin/GrabowskiKR18}}]\label{fact:PRECT}
  \PRECT\ can be solved in $\Oh(m \log^3 m)$ time and $\Oh(m \log^2 m)$ space.
\end{fact}

\subsection{Algorithm for \FFP}
Let us recall that the suffix tree of string $W$, $\ST_W$, is a compacted trie of all the suffixes of~$W$.
It can be computed in $\Oh(n)$ time (see \cite{DBLP:journals/jacm/Farach-ColtonFM00}) and reading the suffixes of $W$ in its preorder traversal yields the suffix array of $W$.
An efficient implementation of \cref{obs:SAint} is known; see~\cite{DBLP:journals/talg/AmirLLS07}.
\begin{lemma}\label{lem:WAQ}
  After $\Oh(n)$-time preprocessing, for a given fragment $u$ of $W$ one can compute in $\Oh(\log \log n)$ time the sets $\Int(u)$ and $\Intr(u)$.
\end{lemma}
\begin{proof}
  It suffices to show how to compute $\Int(u)$.
  For every explicit node of $\ST_W$ we can compute the interval of elements of $\SA_W$ that are located in its subtree.
  This can be done in a bottom-up order in $\Oh(n)$ time.

  A weighted ancestor query in $\ST_W$, given a terminal node $w$ and positive integer $d$, returns the ancestor of $w$ located at depth $d$
  (being an explicit or implicit node).
  Such queries (for any tree of $n$ nodes with positive integer weights of edges) can be answered in $\Oh(\log \log n)$ time after $\Oh(n)$-time preprocessing;
  see~\cite{DBLP:journals/talg/AmirLLS07}.

  A weighted ancestor query can be used to, given a fragment $u$ of $W$, compute the corresponding (explicit or implicit) node $w$ of $\ST_W$.
  The interval stored in the nearest explicit descendant of $w$ equals $\Int(u)$.
\end{proof}

We are now ready to show a solution to \FFP.

\begin{lemma}
  \FFP\ can be solved in $\Oh(n + m (\log \log n + \log^3 m))$ time and $\Oh(n+m \log^2 m)$ space.
\end{lemma}
\begin{proof}
  We construct families $\R_1$ and $\R_2$ of weighted 4-rectangles.
  For every $(x,y) \in \F_1$, we add $\RECT(x,y)$ to $\R_1$ with weight $|x|+|y|$.
  For every $(z,t) \in \F_2$, we add $\RECT'(z,t)$ to $\R_2$ with weight $|z|+|t|$.
  By \cref{obs:red}, the solution to \PRECT\ for $\R_1$ and $\R_2$ is the solution to $\FFP(\F_1,\F_2)$.

  Note that we have $|\R_1|=|\F_1|$ and $|\R_2|=|\F_2|$.
  Using \cref{lem:WAQ} and a linear-time algorithm for constructing $\SA_W$ and $\PA_W$ (and $\SA^{-1}_W$ and $\PA^{-1}_W$)~\cite{DBLP:journals/jacm/Farach-ColtonFM00},
  computation of 4-rectangles $\RECT$, $\RECT'$ can be done in $\Oh(\log \log n)$ time after $\Oh(n)$-time preprocessing.
  This gives $\Oh(n + m \log \log n)$ time in total.
  Finally, \PRECT\ can be solved in $\Oh(m \log^3 m)$ time and $\Oh(m \log^2 m)$ space.
\end{proof}

\noindent As a consequence of all the previous Correctness and Complexity lemmas and the above lemma we obtain the main result.

\thmmain*

\section{Conclusions}
We have presented an $\Oh(n \log^5 n)$-time algorithm for computing the Longest Common Circular Factor (LCCF) of two strings of length $n$.
Let us recall that the Longest Common Factor (LCF) of two strings can be computed in $\Oh(n)$ time.
We leave an open question if the LCCF problem can also be solved in linear time.

\bibliographystyle{plainurl}
\bibliography{lccf}
\end{document}

%% file: rys_synch.tex
\begin{tikzpicture}[scale=0.9]\path (0.0, 0) rectangle node[anchor=south]{\tt b} ++(0.5, 0.5);
\path (0.5, 0) rectangle node[anchor=south]{\tt b} ++(0.5, 0.5);
\path (1.0, 0) rectangle node[anchor=south]{\tt a} ++(0.5, 0.5);
\path (1.5, 0) rectangle node[anchor=south]{\tt b} ++(0.5, 0.5);
\path (2.0, 0) rectangle node[anchor=south]{\tt a} ++(0.5, 0.5);
\path (2.5, 0) rectangle node[anchor=south]{\tt b} ++(0.5, 0.5);
\path (3.0, 0) rectangle node[anchor=south]{\tt a} ++(0.5, 0.5);
\path (3.5, 0) rectangle node[anchor=south]{\tt a} ++(0.5, 0.5);
\path (4.0, 0) rectangle node[anchor=south]{\tt a} ++(0.5, 0.5);
\path (4.5, 0) rectangle node[anchor=south]{\tt a} ++(0.5, 0.5);
\path (5.0, 0) rectangle node[anchor=south]{\tt a} ++(0.5, 0.5);
\path (5.5, 0) rectangle node[anchor=south]{\tt a} ++(0.5, 0.5);
\path (6.0, 0) rectangle node[anchor=south]{\tt a} ++(0.5, 0.5);
\path (6.5, 0) rectangle node[anchor=south]{\tt a} ++(0.5, 0.5);
\path (7.0, 0) rectangle node[anchor=south]{\tt b} ++(0.5, 0.5);
\path (7.5, 0) rectangle node[anchor=south]{\tt b} ++(0.5, 0.5);
\path (8.0, 0) rectangle node[anchor=south]{\tt a} ++(0.5, 0.5);
\path (8.5, 0) rectangle node[anchor=south]{\tt b} ++(0.5, 0.5);
\path (9.0, 0) rectangle node[anchor=south]{\tt a} ++(0.5, 0.5);
\path (9.5, 0) rectangle node[anchor=south]{\tt b} ++(0.5, 0.5);
\path (10.0, 0) rectangle node[anchor=south]{\tt a} ++(0.5, 0.5);
\path (10.5, 0) rectangle node[anchor=south]{\tt b} ++(0.5, 0.5);
\path (11.0, 0) rectangle node[anchor=south]{\tt a} ++(0.5, 0.5);
\path (11.5, 0) rectangle node[anchor=south]{\tt b} ++(0.5, 0.5);
\definecolor{black}{gray}{0.95};
\definecolor{red}{gray}{0.7999999999999999};
\definecolor{blue}{gray}{0.6499999999999999};
\definecolor{green}{gray}{0.4999999999999999};
\draw [fill=none] (0.0, 1.0) rectangle ++(3.5, 0.25);
\draw [fill=none] (0.5, 1.25) rectangle ++(3.5, 0.25);
\draw [fill=black] (1.0, 1.0) rectangle ++(2.0, 0.25);
\draw [fill=black] (1.0, 1.25) rectangle ++(2.0, 0.25);
\draw [fill=none] (1.0, -0.25) rectangle ++(3.5, 0.25);
\draw [fill=red] (2.5, -0.25) rectangle ++(2.0, 0.25);
\draw [fill=none] (1.5, -0.5) rectangle ++(3.5, 0.25);
\draw [fill=red] (2.5, -0.5) rectangle ++(2.0, 0.25);
\draw [fill=none] (2.0, -0.75) rectangle ++(3.5, 0.25);
\draw [fill=red] (2.5, -0.75) rectangle ++(2.0, 0.25);
\draw [fill=none] (2.5, -1.0) rectangle ++(3.5, 0.25);
\draw [fill=red] (2.5, -1.0) rectangle ++(2.0, 0.25);
\draw [fill=none] (4.0, 1.0) rectangle ++(3.5, 0.25);
\draw [fill=blue] (5.5, 1.0) rectangle ++(2.0, 0.25);
\draw [fill=none] (4.5, 1.25) rectangle ++(3.5, 0.25);
\draw [fill=blue] (5.5, 1.25) rectangle ++(2.0, 0.25);
\draw [fill=none] (5.0, 1.5) rectangle ++(3.5, 0.25);
\draw [fill=blue] (5.5, 1.5) rectangle ++(2.0, 0.25);
\draw [fill=none] (5.5, 1.75) rectangle ++(3.5, 0.25);
\draw [fill=blue] (5.5, 1.75) rectangle ++(2.0, 0.25);
\draw [fill=none] (6.0, -0.25) rectangle ++(3.5, 0.25);
\draw [fill=none] (6.5, -0.5) rectangle ++(3.5, 0.25);
\draw [fill=none] (7.0, -0.75) rectangle ++(3.5, 0.25);
\draw [fill=none] (7.5, -1.0) rectangle ++(3.5, 0.25);
\draw [fill=none] (8.0, -1.25) rectangle ++(3.5, 0.25);
\draw [fill=green] (7.0, -0.25) rectangle ++(2.0, 0.25);
\draw [fill=black] (8.0, -0.5) rectangle ++(2.0, 0.25);
\draw [fill=black] (8.0, -0.75) rectangle ++(2.0, 0.25);
\draw [fill=black] (8.0, -1.0) rectangle ++(2.0, 0.25);
\draw [fill=black] (8.0, -1.25) rectangle ++(2.0, 0.25);
\draw [fill=none] (8.5, 1.0) rectangle ++(3.5, 0.25);
\draw [fill=black] (9.0, 1.0) rectangle ++(2.0, 0.25);
\draw[densely dotted] (1.0, 0.25) -- ++(0, 1.0);
\draw[densely dotted] (3.0, 0.25) -- ++(0, 1.0);
\draw[densely dotted] (5.5, 0.25) -- ++(0, 1.0);
\draw[densely dotted] (7.5, 0.25) -- ++(0, 1.0);
\draw[densely dotted] (9.0, 0.25) -- ++(0, 1.0);
\draw[densely dotted] (11.0, 0.25) -- ++(0, 1.0);
\draw[densely dotted] (2.5, 0.75) -- ++(0, -1.0);
\draw[densely dotted] (4.5, 0.75) -- ++(0, -1.0);
\draw[densely dotted] (7.0, 0.75) -- ++(0, -1.0);
\draw[densely dotted] (9.0, 0.75) -- ++(0, -1.0);
\end{tikzpicture}

%% file: _fig_concrete_ex.tex
\begin{tikzpicture}[scale=0.8]
  \foreach \x/\c in {-0.5/a,-0.25/d,0/b,0.25/a, 0.6/b,0.85/a,1.1/a, 1.45/a,1.7/b,1.95/a, 2.3/b,2.55/a,2.8/c, 3.15/b,3.4/c, 3.75/b,4.0/a,4.25/a,4.5/c,4.75/c}{
    \draw (\x,0) node[above] {\tt \c};
  }
  \draw (-1.5,0) node[above] {$S$};
  \draw (0.5,0) -- node[below] {$\vphantom{yt}t$} (1.2,0);
  \draw (1.35,0.6) -- (1.35,0.8) -- node[above] {$\vphantom{yt}x$} (2.05,0.8) -- (2.05,0.6);
  \draw (2.2,0.6) -- (2.2,0.8) -- node[above] {$\vphantom{yt}y$} (2.9,0.8) -- (2.9,0.6);
  \draw (3.05,0) -- node[below] {$\vphantom{yt}z$} (3.5,0);
\begin{scope}[xshift=8cm]
  \foreach \x/\c in {-0.5/b,-0.25/a,0/a,0.25/c, 0.6/b,0.85/a,1.1/c, 1.45/b,1.7/c, 2.05/b,2.3/a,2.55/a, 2.9/a,3.15/b,3.4/a, 3.75/b,4.0/a,4.25/c,4.5/c,4.75/a}{
    \draw (\x,0) node[above] {\tt \c};
  }
  \draw (-1.5,0) node[above] {$T$};
  \draw (0.5,0) -- node[below] {$\vphantom{yt}y$} (1.2,0);
  \draw (1.35,0.6) -- (1.35,0.8) -- node[above] {$\vphantom{yt}z$} (1.8,0.8) -- (1.8,0.6);
  \draw (1.95,0.6) -- (1.95,0.8) -- node[above] {$\vphantom{yt}t$} (2.65,0.8) -- (2.65,0.6);
  \draw (2.8,0) -- node[below] {$\vphantom{yt}x$} (3.5,0);
\end{scope}
\end{tikzpicture}

%% file: _fig_nonperiod_image.tex
\begin{tikzpicture}
\draw (-1,0) node {$S$};
\draw[decoration={brace, raise=1.55cm, amplitude=10pt}, decorate] (0, 0) -- (4.0, 0) node[black,midway,yshift=2.1cm] {$2^{a+2}-2$};
\draw[decoration={brace, raise=1.55cm, amplitude=10pt}, decorate] (4.0, 0) -- (12.0, 0) node[black,midway,yshift=2.1cm] {$2^{b + 2}-2$};
\node at (0, 0) {\textbullet};
\node at (1.5, 0) {\textbullet};
\draw (0, 0) -- node[above=5pt, yshift=1cm]{} (1.5, 0);
\draw[dashed] (0, 10pt) -- (0, 40pt);
\node at (1.5, 0) {\textbullet};
\node at (4.0, 0) {\textbullet};
\draw (1.5, 0) -- node[above=5pt, yshift=.9cm]{$x$} (4.0, 0);
\draw[pattern=north west lines] (1.5, 0) rectangle node[above, yshift=1pt] {$\vphantom{\alpha\beta'}\alpha$} (2.5, 6pt);
\draw[dashed] (1.5, 26pt) -- (1.5, 40pt);
\node at (4.0, 0) {\textbullet};
\node at (7.0, 0) {\textbullet};
\draw (4.0, 0) -- node[above=5pt, yshift=.9cm]{$y$} (7.0, 0);
\draw[dashed] (4.0, 26pt) -- (4.0, 40pt);
\node at (12.0, 0) {\textbullet};
\node at (7.0, 0) {\textbullet};
\draw (7.0, 0) -- node[above=5pt, yshift=1cm]{} (12.0, 0);
\draw[pattern=north west lines] (7.0, 0) rectangle node[above, yshift=1pt] {$\vphantom{\alpha\beta'}\beta$} (9.0, 6pt);
\draw[dashed] (7.0, 26pt) -- (7.0, 40pt);
\draw[rounded corners=5pt] (1.0, 0) rectangle node[above, yshift=10pt] {$U$} (4.0, 20pt);
\draw[rounded corners=5pt] (11.5, 0) rectangle node[above, yshift=10pt] {$V$} (4.0, 20pt);
\draw[dashed] (12.0, 10pt) -- (12.0, 40pt);
\node[below=4pt] at (4.0, 0) {\large $i$};
\end{tikzpicture}

%% file: _fig_nonperiod_image2.tex
\begin{tikzpicture}
\draw (-1,0) node {$T$};
\draw[decoration={brace, raise=1.55cm, amplitude=10pt}, decorate] (0, 0) -- (8.0, 0) node[black,midway,yshift=2.1cm] {$2^{b+2}-2$};
\draw[decoration={brace, raise=1.55cm, amplitude=10pt}, decorate] (8.0, 0) -- (12.0, 0) node[black,midway,yshift=2.1cm] {$2^{a + 2}-2$};
\node at (0, 0) {\textbullet};
\node at (3.5, 0) {\textbullet};
\draw (0, 0) -- node[above=5pt, yshift=1cm]{} (3.5, 0);
\draw[dashed] (0, 10pt) -- (0, 40pt);
\node at (3.5, 0) {\textbullet};
\node at (8.0, 0) {\textbullet};
\draw (3.5, 0) -- node[above=5pt, yshift=.9cm]{$z$} (8.0, 0);
\draw[pattern=north west lines] (3.5, 0) rectangle node[above, yshift=1pt] {$\vphantom{\alpha\beta'}\beta'$} (5.5, 6pt);
\draw[dashed] (3.5, 26pt) -- (3.5, 40pt);
\node at (8.0, 0) {\textbullet};
\node at (8.5, 0) {\textbullet};
\draw (8.0, 0) -- node[above=5pt, yshift=.9cm]{$t$} (8.5, 0);
\draw[dashed] (8.0, 26pt) -- (8.0, 40pt);
\node at (8.5, 0) {\textbullet};
\node at (12.0, 0) {\textbullet};
\draw (8.5, 0) -- node[above=5pt, yshift=1cm]{} (12.0, 0);
\draw[pattern=north west lines] (8.5, 0) rectangle node[above, yshift=1pt] {$\vphantom{\alpha\beta'}\alpha'$} (9.5, 6pt);
\draw[dashed] (8.5, 26pt) -- (8.5, 40pt);
\draw[rounded corners=5pt] (0.5, 0) rectangle node[above, yshift=10pt] {$V$} (8.0, 20pt);
\draw[rounded corners=5pt] (11.0, 0) rectangle node[above, yshift=10pt] {$U$} (8.0, 20pt);
\draw[dashed] (12.0, 10pt) -- (12.0, 40pt);
\node[below=4pt] at (8.0, 0) {\large $i'$};
\end{tikzpicture}

%% file: _rys1.tex
\begin{tikzpicture}[scale=0.8]

\draw (0, 0) node[left]{$W$} -- (6.25, 0);

\begin{scope}

\clip (1.3333333333333335, -1) rectangle (6.25, 1);

\draw (0.8333333333333334, 0) to[out=45, in=180] (1.25, 0.55) node[above]{} to[out=0, in=135] (1.6666666666666667, 0);

\draw (1.6666666666666667, 0) to[out=45, in=180] (2.0833333333333335, 0.5499999999999999) node[above]{$\lambda$} to[out=0, in=135] (2.5, 0);

\draw (2.5, 0) to[out=45, in=180] (2.916666666666667, 0.5500000000000002) node[above]{} to[out=0, in=135] (3.3333333333333335, 0);

\draw (3.3333333333333335, 0) to[out=45, in=180] (3.75, 0.5500000000000002) node[above]{} to[out=0, in=135] (4.166666666666667, 0);

\draw (4.166666666666667, 0) to[out=45, in=180] (4.583333333333334, 0.5499999999999998) node[above]{} to[out=0, in=135] (5.0, 0);

\draw (5.0, 0) to[out=45, in=180] (5.416666666666666, 0.5499999999999998) node[above]{} to[out=0, in=135] (5.833333333333333, 0);

\draw (5.833333333333333, 0) to[out=45, in=180] (6.25, 0.5499999999999998) node[above]{} to[out=0, in=135] (6.666666666666666, 0);

\draw (6.666666666666666, 0) to[out=45, in=180] (7.083333333333332, 0.5499999999999998) node[above]{} to[out=0, in=135] (7.499999999999999, 0);

\end{scope}




\draw (2.666666666666667, -1) rectangle ++(3.5833333333333335, 0.5) node[pos=.5]{};

\draw[-latex] (2.666666666666667, -0.75) -- node[above]{$|\lambda|$} ++(-0.8333333333333334, 0);

\draw(1.25, 0.5) node{*};

\end{tikzpicture}

%% file: _rys2.tex
\begin{tikzpicture}[scale=0.8]
\draw (2.666666666666667, 0) node[left]{$W'$} -- (10, 0);
\begin{scope}
\clip (2.666666666666667, -1) rectangle (9.0, 1);
\draw (2.5, 0) to[out=45, in=180] (2.916666666666667, 0.5500000000000002) node[above]{} to[out=0, in=135] (3.3333333333333335, 0);
\draw (3.3333333333333335, 0) to[out=45, in=180] (3.75, 0.5500000000000002) node[above]{$\lambda$} to[out=0, in=135] (4.166666666666667, 0);
\draw (4.166666666666667, 0) to[out=45, in=180] (4.583333333333334, 0.5499999999999998) node[above]{} to[out=0, in=135] (5.0, 0);
\draw (5.0, 0) to[out=45, in=180] (5.416666666666666, 0.5499999999999998) node[above]{} to[out=0, in=135] (5.833333333333333, 0);
\draw (5.833333333333333, 0) to[out=45, in=180] (6.25, 0.5499999999999998) node[above]{} to[out=0, in=135] (6.666666666666666, 0);
\draw (6.666666666666666, 0) to[out=45, in=180] (7.083333333333332, 0.5499999999999998) node[above]{} to[out=0, in=135] (7.499999999999999, 0);
\draw (7.499999999999999, 0) to[out=45, in=180] (7.916666666666666, 0.5499999999999998) node[above]{} to[out=0, in=135] (8.333333333333332, 0);
\draw (8.333333333333332, 0) to[out=45, in=180] (8.75, 0.5500000000000004) node[above]{} to[out=0, in=135] (9.166666666666666, 0);
\draw (9.166666666666666, 0) to[out=45, in=180] (9.583333333333332, 0.5500000000000004) node[above]{} to[out=0, in=135] (10.0, 0);
\end{scope}
\draw (2.666666666666667, -1) rectangle ++(3.5833333333333335, 0.5) node[pos=.5]{};
\draw[-latex] (6.25, -0.75) -- node[above]{$|\lambda|$} ++(0.8333333333333334, 0);
\end{tikzpicture}

%% file: _rys11.tex
\begin{tikzpicture}[scale=1.2]
\draw (0, 0) -- (10, 0);
\begin{scope}
\clip (0.10416666666666666, -1) rectangle (3.333333333333334, 1.5);
\draw (0.0, 0) to[out=45, in=180] (0.1736111111111111, 0.22916666666666666) node[above]{} to[out=0, in=135] (0.3472222222222222, 0);
\draw (0.3472222222222222, 0) to[out=45, in=180] (0.5208333333333333, 0.22916666666666666) node[above]{$\alpha$} to[out=0, in=135] (0.6944444444444444, 0);
\draw (0.3472222222222222, 0) -- node[below]{$x_1$} (0.6944444444444444, 0);
\draw (0.6944444444444444, 0) to[out=45, in=180] (0.8680555555555555, 0.2291666666666666) node[above]{$\alpha$} to[out=0, in=135] (1.0416666666666665, 0);
\draw (0.6944444444444444, 0) -- node[below]{$x_2$} (1.0416666666666665, 0);
\draw (1.0416666666666665, 0) to[out=45, in=180] (1.2152777777777777, 0.22916666666666674) node[above]{} to[out=0, in=135] (1.3888888888888888, 0);
\draw (1.3888888888888888, 0) to[out=45, in=180] (1.5625, 0.22916666666666674) node[above]{} to[out=0, in=135] (1.7361111111111112, 0);
\draw (1.7361111111111112, 0) to[out=45, in=180] (1.9097222222222223, 0.22916666666666674) node[above]{} to[out=0, in=135] (2.0833333333333335, 0);
\draw (2.0833333333333335, 0) to[out=45, in=180] (2.2569444444444446, 0.22916666666666674) node[above]{} to[out=0, in=135] (2.430555555555556, 0);
\draw (2.430555555555556, 0) to[out=45, in=180] (2.604166666666667, 0.22916666666666674) node[above]{} to[out=0, in=135] (2.777777777777778, 0);
\draw (2.777777777777778, 0) to[out=45, in=180] (2.9513888888888893, 0.22916666666666674) node[above]{$\alpha$} to[out=0, in=135] (3.1250000000000004, 0);
\draw (2.777777777777778, 0) -- node[below]{$x_3$} (3.1250000000000004, 0);
\draw (3.1250000000000004, 0) to[out=45, in=180] (3.2986111111111116, 0.22916666666666674) node[above]{} to[out=0, in=135] (3.4722222222222228, 0);
\draw (3.4722222222222228, 0) to[out=45, in=180] (3.645833333333334, 0.22916666666666674) node[above]{} to[out=0, in=135] (3.819444444444445, 0);
\end{scope}
\begin{scope}
\clip (3.3333333333333335, -1) rectangle (10.0, 1.5);
\draw (3.0303030303030303, 0) to[out=45, in=180] (3.5353535353535355, 0.6666666666666669) node[above]{} to[out=0, in=135] (4.040404040404041, 0);
\draw (4.040404040404041, 0) to[out=45, in=180] (4.545454545454546, 0.6666666666666669) node[above]{$\beta$} to[out=0, in=135] (5.050505050505051, 0);
\draw (4.040404040404041, 0) -- node[below]{$y_1$} (5.050505050505051, 0);
\draw (5.050505050505051, 0) to[out=45, in=180] (5.555555555555556, 0.6666666666666669) node[above]{} to[out=0, in=135] (6.060606060606061, 0);
\draw (6.060606060606061, 0) to[out=45, in=180] (6.565656565656567, 0.6666666666666669) node[above]{} to[out=0, in=135] (7.070707070707072, 0);
\draw (7.070707070707072, 0) to[out=45, in=180] (7.575757575757576, 0.6666666666666663) node[above]{$\beta$} to[out=0, in=135] (8.080808080808081, 0);
\draw (7.070707070707072, 0) -- node[below]{$y_2$} (8.080808080808081, 0);
\draw (8.080808080808081, 0) to[out=45, in=180] (8.585858585858587, 0.6666666666666669) node[above]{$\beta$} to[out=0, in=135] (9.090909090909092, 0);
\draw (8.080808080808081, 0) -- node[below]{$y_3$} (9.090909090909092, 0);
\draw (9.090909090909092, 0) to[out=45, in=180] (9.595959595959597, 0.6666666666666669) node[above]{} to[out=0, in=135] (10.101010101010102, 0);
\draw (10.101010101010102, 0) to[out=45, in=180] (10.606060606060607, 0.6666666666666669) node[above]{} to[out=0, in=135] (11.111111111111112, 0);
\end{scope}
\draw[dashed] (3.3333333333333335, -0.1) -- (3.3333333333333335, 1) node[above]{$i$};
\draw[latex-latex](3.3333333333333335, 1.5) -- node[above]{$2^{b + 1}-1$} (6.5, 1.5);
\draw[latex-latex](3.3333333333333335, 2) -- node[above]{$2^{b + 2}-2$} (9.666666666666666, 2);
\draw[latex-latex](1.7500000000000002, 1.5) -- node[above]{$2^{a + 1}-1$} (3.3333333333333335, 1.5);
\draw[latex-latex](0.16666666666666696, 2) -- node[above]{$2^{a + 2}-2$} (3.3333333333333335, 2);
\end{tikzpicture}

%% file: rys3.tex
\begin{tikzpicture}[scale=0.5]
\useasboundingbox (0,-1) rectangle (10, 4);\draw (0, 0) -- (10, 0);
\begin{scope}
\begin{pgfinterruptboundingbox}
\clip (2.0833333333333335, -1.5) rectangle (6.9444444444444455, 1.5);
\end{pgfinterruptboundingbox}
\draw (1.6666666666666667, 0) to[out=45, in=180] (2.0833333333333335, 0.5499999999999999) node[above]{} to[out=0, in=135] (2.5, 0);
\draw (2.5, 0) to[out=45, in=180] (2.916666666666667, 0.5500000000000002) node[above]{\footnotesize $\lambda$} to[out=0, in=135] (3.3333333333333335, 0);
\draw[thick] (3.75, 0) ellipse (0.4166666666666667 and 0.2);
\draw (3.3333333333333335, 0) to[out=45, in=180] (3.75, 0.5500000000000002) node[above]{} to[out=0, in=135] (4.166666666666667, 0);
\draw (4.166666666666667, 0) to[out=45, in=180] (4.583333333333334, 0.5499999999999998) node[above]{} to[out=0, in=135] (5.0, 0);
\draw (5.0, 0) to[out=45, in=180] (5.416666666666666, 0.5499999999999998) node[above]{} to[out=0, in=135] (5.833333333333333, 0);
\draw (5.833333333333333, 0) to[out=45, in=180] (6.25, 0.5499999999999998) node[above]{} to[out=0, in=135] (6.666666666666666, 0);
\draw (6.666666666666666, 0) to[out=45, in=180] (7.083333333333332, 0.5499999999999998) node[above]{} to[out=0, in=135] (7.499999999999999, 0);
\draw (7.499999999999999, 0) to[out=45, in=180] (7.916666666666666, 0.5499999999999998) node[above]{} to[out=0, in=135] (8.333333333333332, 0);
\end{scope}
\draw[dashed] (6.9444444444444455, -0.1) -- (6.9444444444444455, 0.75) node[above]{\footnotesize $i$};
\draw[latex-latex](1.3888888888888884, 1.6) -- node[above=-1pt]{\footnotesize $2^{a + 2}-2$} (6.9444444444444455, 1.6);
\draw[latex-latex](4.166666666666667, 2.5) -- node[above=-1pt]{\footnotesize $2^{a + 1}-1$} (6.9444444444444455, 2.5);
\begin{scope}
\draw (2.666666666666667, -1) rectangle ++(4.277777777777779, 0.5) node[pos=.5]{};
\path[clip]
(0.0, -1.5)
to (0.0, 0) to [out=45, in=180] (0.4166666666666667, 0.55) to[out=0, in=135] (0.8333333333333334, 0)
to (0.8333333333333334, 0) to [out=45, in=180] (1.25, 0.55) to[out=0, in=135] (1.6666666666666667, 0)
to (1.6666666666666667, 0) to [out=45, in=180] (2.0833333333333335, 0.5499999999999999) to[out=0, in=135] (2.5, 0)
to (2.5, 0) to [out=45, in=180] (2.916666666666667, 0.5500000000000002) to[out=0, in=135] (3.3333333333333335, 0)
to (3.3333333333333335, 0) to [out=45, in=180] (3.75, 0.5500000000000002) to[out=0, in=135] (4.166666666666667, 0)
to (4.166666666666667, 0) to [out=45, in=180] (4.583333333333334, 0.5499999999999998) to[out=0, in=135] (5.0, 0)
to (5.0, 0) to [out=45, in=180] (5.416666666666666, 0.5499999999999998) to[out=0, in=135] (5.833333333333333, 0)
to (5.833333333333333, 0) to [out=45, in=180] (6.25, 0.5499999999999998) to[out=0, in=135] (6.666666666666666, 0)
to (6.666666666666666, 0) to [out=45, in=180] (7.083333333333332, 0.5499999999999998) to[out=0, in=135] (7.499999999999999, 0)
to (7.499999999999999, 0) to [out=45, in=180] (7.916666666666666, 0.5499999999999998) to[out=0, in=135] (8.333333333333332, 0)
to (8.333333333333332, 0) to [out=45, in=180] (8.75, 0.5500000000000004) to[out=0, in=135] (9.166666666666666, 0)
to (9.166666666666666, 0) to [out=45, in=180] (9.583333333333332, 0.5500000000000004) to[out=0, in=135] (10.0, 0)
to (10.0, 0) to [out=45, in=180] (10.416666666666668, 0.5500000000000004) to[out=0, in=135] (10.833333333333334, 0)
to (10.833333333333334, 0) to [out=45, in=180] (11.25, 0.5500000000000004) to[out=0, in=135] (11.666666666666668, 0)
to (11.666666666666668, -1.5)
to (0.0, -1.5)
;\draw[densely dotted] (2.666666666666667, -0.5) -- (2.666666666666667, 1.5);
\end{scope}
\draw(1.9833333333333334, 0.5) node{*};
\end{tikzpicture}

%% file: rys4.tex
\begin{tikzpicture}[scale=0.5]
\useasboundingbox (0,-1) rectangle (10, 4);\draw (0, 0) -- (10, 0);
\begin{scope}
\begin{pgfinterruptboundingbox}
\clip (1.8333333333333335, -1.5) rectangle (8.166666666666666, 1.5);
\end{pgfinterruptboundingbox}
\draw (1.6666666666666667, 0) to[out=45, in=180] (2.0833333333333335, 0.5499999999999999) node[above]{} to[out=0, in=135] (2.5, 0);
\draw[thick] (2.9166666666666665, 0) ellipse (0.4166666666666667 and 0.2);
\draw (2.5, 0) to[out=45, in=180] (2.916666666666667, 0.5500000000000002) node[above]{\footnotesize $\lambda$} to[out=0, in=135] (3.3333333333333335, 0);
\draw (3.3333333333333335, 0) to[out=45, in=180] (3.75, 0.5500000000000002) node[above]{} to[out=0, in=135] (4.166666666666667, 0);
\draw (4.166666666666667, 0) to[out=45, in=180] (4.583333333333334, 0.5499999999999998) node[above]{} to[out=0, in=135] (5.0, 0);
\draw (5.0, 0) to[out=45, in=180] (5.416666666666666, 0.5499999999999998) node[above]{} to[out=0, in=135] (5.833333333333333, 0);
\draw (5.833333333333333, 0) to[out=45, in=180] (6.25, 0.5499999999999998) node[above]{} to[out=0, in=135] (6.666666666666666, 0);
\draw (6.666666666666666, 0) to[out=45, in=180] (7.083333333333332, 0.5499999999999998) node[above]{} to[out=0, in=135] (7.499999999999999, 0);
\draw (7.499999999999999, 0) to[out=45, in=180] (7.916666666666666, 0.5499999999999998) node[above]{} to[out=0, in=135] (8.333333333333332, 0);
\draw (8.333333333333332, 0) to[out=45, in=180] (8.75, 0.5500000000000004) node[above]{} to[out=0, in=135] (9.166666666666666, 0);
\end{scope}
\draw[dashed] (1.8333333333333335, -0.1) -- (1.8333333333333335, 0.75) node[above]{\footnotesize $j$};
\draw[latex-latex](1.8333333333333335, 1.6) -- node[above=-1pt]{\footnotesize $2^{a + 2}-2$} (8.166666666666666, 1.6);
\draw[latex-latex](1.8333333333333335, 2.5) -- node[above=-1pt]{\footnotesize $2^{a + 1}-1$} (5.0, 2.5);
\begin{scope}
\draw (1.8333333333333335, -1) rectangle ++(4.277777777777779, 0.5) node[pos=.5]{};
\path[clip]
(0.0, -1.5)
to (0.0, 0) to [out=45, in=180] (0.4166666666666667, 0.55) to[out=0, in=135] (0.8333333333333334, 0)
to (0.8333333333333334, 0) to [out=45, in=180] (1.25, 0.55) to[out=0, in=135] (1.6666666666666667, 0)
to (1.6666666666666667, 0) to [out=45, in=180] (2.0833333333333335, 0.5499999999999999) to[out=0, in=135] (2.5, 0)
to (2.5, 0) to [out=45, in=180] (2.916666666666667, 0.5500000000000002) to[out=0, in=135] (3.3333333333333335, 0)
to (3.3333333333333335, 0) to [out=45, in=180] (3.75, 0.5500000000000002) to[out=0, in=135] (4.166666666666667, 0)
to (4.166666666666667, 0) to [out=45, in=180] (4.583333333333334, 0.5499999999999998) to[out=0, in=135] (5.0, 0)
to (5.0, 0) to [out=45, in=180] (5.416666666666666, 0.5499999999999998) to[out=0, in=135] (5.833333333333333, 0)
to (5.833333333333333, 0) to [out=45, in=180] (6.25, 0.5499999999999998) to[out=0, in=135] (6.666666666666666, 0)
to (6.666666666666666, 0) to [out=45, in=180] (7.083333333333332, 0.5499999999999998) to[out=0, in=135] (7.499999999999999, 0)
to (7.499999999999999, 0) to [out=45, in=180] (7.916666666666666, 0.5499999999999998) to[out=0, in=135] (8.333333333333332, 0)
to (8.333333333333332, 0) to [out=45, in=180] (8.75, 0.5500000000000004) to[out=0, in=135] (9.166666666666666, 0)
to (9.166666666666666, 0) to [out=45, in=180] (9.583333333333332, 0.5500000000000004) to[out=0, in=135] (10.0, 0)
to (10.0, 0) to [out=45, in=180] (10.416666666666668, 0.5500000000000004) to[out=0, in=135] (10.833333333333334, 0)
to (10.833333333333334, 0) to [out=45, in=180] (11.25, 0.5500000000000004) to[out=0, in=135] (11.666666666666668, 0)
to (11.666666666666668, -1.5)
to (0.0, -1.5)
;\draw[densely dotted] (6.1111111111111125, -0.5) -- (6.1111111111111125, 1.5);
\end{scope}
\end{tikzpicture}

%% file: rys5.tex
\begin{tikzpicture}[scale=0.5]
\useasboundingbox (0,-1) rectangle (10, 4);\draw (0, 0) -- (10, 0);
\begin{scope}
\begin{pgfinterruptboundingbox}
\clip (0.4166666666666667, -1.5) rectangle (5.416666666666667, 1.5);
\end{pgfinterruptboundingbox}
\draw (0.0, 0) to[out=45, in=180] (0.4166666666666667, 0.55) node[above]{} to[out=0, in=135] (0.8333333333333334, 0);
\draw[thick] (1.25, 0) ellipse (0.4166666666666667 and 0.2);
\draw (0.8333333333333334, 0) to[out=45, in=180] (1.25, 0.55) node[above]{} to[out=0, in=135] (1.6666666666666667, 0);
\draw (1.6666666666666667, 0) to[out=45, in=180] (2.0833333333333335, 0.5499999999999999) node[above]{} to[out=0, in=135] (2.5, 0);
\draw (2.5, 0) to[out=45, in=180] (2.916666666666667, 0.5500000000000002) node[above]{} to[out=0, in=135] (3.3333333333333335, 0);
\draw (3.3333333333333335, 0) to[out=45, in=180] (3.75, 0.5500000000000002) node[above]{} to[out=0, in=135] (4.166666666666667, 0);
\draw (4.166666666666667, 0) to[out=45, in=180] (4.583333333333334, 0.5499999999999998) node[above]{} to[out=0, in=135] (5.0, 0);
\draw (5.0, 0) to[out=45, in=180] (5.416666666666666, 0.5499999999999998) node[above]{} to[out=0, in=135] (5.833333333333333, 0);
\draw (5.833333333333333, 0) to[out=45, in=180] (6.25, 0.5499999999999998) node[above]{} to[out=0, in=135] (6.666666666666666, 0);
\end{scope}
\draw[dashed] (5.416666666666667, -0.1) -- (5.416666666666667, 0.75) node[above]{\footnotesize $i$};
\begin{scope}
\draw (0.4166666666666667, -1) rectangle ++(5.0, 0.5) node[pos=.5]{};
\path[clip]
(0.0, -1.5)
to (0.0, 0) to [out=45, in=180] (0.4166666666666667, 0.55) to[out=0, in=135] (0.8333333333333334, 0)
to (0.8333333333333334, 0) to [out=45, in=180] (1.25, 0.55) to[out=0, in=135] (1.6666666666666667, 0)
to (1.6666666666666667, 0) to [out=45, in=180] (2.0833333333333335, 0.5499999999999999) to[out=0, in=135] (2.5, 0)
to (2.5, 0) to [out=45, in=180] (2.916666666666667, 0.5500000000000002) to[out=0, in=135] (3.3333333333333335, 0)
to (3.3333333333333335, 0) to [out=45, in=180] (3.75, 0.5500000000000002) to[out=0, in=135] (4.166666666666667, 0)
to (4.166666666666667, 0) to [out=45, in=180] (4.583333333333334, 0.5499999999999998) to[out=0, in=135] (5.0, 0)
to (5.0, 0) to [out=45, in=180] (5.416666666666666, 0.5499999999999998) to[out=0, in=135] (5.833333333333333, 0)
to (5.833333333333333, 0) to [out=45, in=180] (6.25, 0.5499999999999998) to[out=0, in=135] (6.666666666666666, 0)
to (6.666666666666666, 0) to [out=45, in=180] (7.083333333333332, 0.5499999999999998) to[out=0, in=135] (7.499999999999999, 0)
to (7.499999999999999, 0) to [out=45, in=180] (7.916666666666666, 0.5499999999999998) to[out=0, in=135] (8.333333333333332, 0)
to (8.333333333333332, 0) to [out=45, in=180] (8.75, 0.5500000000000004) to[out=0, in=135] (9.166666666666666, 0)
to (9.166666666666666, 0) to [out=45, in=180] (9.583333333333332, 0.5500000000000004) to[out=0, in=135] (10.0, 0)
to (10.0, 0) to [out=45, in=180] (10.416666666666668, 0.5500000000000004) to[out=0, in=135] (10.833333333333334, 0)
to (10.833333333333334, 0) to [out=45, in=180] (11.25, 0.5500000000000004) to[out=0, in=135] (11.666666666666668, 0)
to (11.666666666666668, -1.5)
to (0.0, -1.5)
;\draw[densely dotted] (0.4166666666666667, -0.5) -- (0.4166666666666667, 1.5);
\end{scope}
\end{tikzpicture}

%% file: rys6.tex
\begin{tikzpicture}[scale=0.5]
\useasboundingbox (0,-1) rectangle (10, 4);\draw (0, 0) -- (10, 0);
\begin{scope}
\begin{pgfinterruptboundingbox}
\clip (2.0833333333333335, -1.5) rectangle (7.916666666666667, 1.5);
\end{pgfinterruptboundingbox}
\draw (1.6666666666666667, 0) to[out=45, in=180] (2.0833333333333335, 0.5499999999999999) node[above]{} to[out=0, in=135] (2.5, 0);
\draw[thick] (2.9166666666666665, 0) ellipse (0.4166666666666667 and 0.2);
\draw (2.5, 0) to[out=45, in=180] (2.916666666666667, 0.5500000000000002) node[above]{} to[out=0, in=135] (3.3333333333333335, 0);
\draw (3.3333333333333335, 0) to[out=45, in=180] (3.75, 0.5500000000000002) node[above]{} to[out=0, in=135] (4.166666666666667, 0);
\draw (4.166666666666667, 0) to[out=45, in=180] (4.583333333333334, 0.5499999999999998) node[above]{} to[out=0, in=135] (5.0, 0);
\draw (5.0, 0) to[out=45, in=180] (5.416666666666666, 0.5499999999999998) node[above]{} to[out=0, in=135] (5.833333333333333, 0);
\draw (5.833333333333333, 0) to[out=45, in=180] (6.25, 0.5499999999999998) node[above]{} to[out=0, in=135] (6.666666666666666, 0);
\draw (6.666666666666666, 0) to[out=45, in=180] (7.083333333333332, 0.5499999999999998) node[above]{} to[out=0, in=135] (7.499999999999999, 0);
\draw (7.499999999999999, 0) to[out=45, in=180] (7.916666666666666, 0.5499999999999998) node[above]{} to[out=0, in=135] (8.333333333333332, 0);
\draw (8.333333333333332, 0) to[out=45, in=180] (8.75, 0.5500000000000004) node[above]{} to[out=0, in=135] (9.166666666666666, 0);
\end{scope}
\draw[dashed] (2.0833333333333335, -0.1) -- (2.0833333333333335, 0.75) node[above]{\footnotesize $j$};
\begin{scope}
\draw (2.0833333333333335, -1) rectangle ++(5.0, 0.5) node[pos=.5]{};
\path[clip]
(0.0, -1.5)
to (0.0, 0) to [out=45, in=180] (0.4166666666666667, 0.55) to[out=0, in=135] (0.8333333333333334, 0)
to (0.8333333333333334, 0) to [out=45, in=180] (1.25, 0.55) to[out=0, in=135] (1.6666666666666667, 0)
to (1.6666666666666667, 0) to [out=45, in=180] (2.0833333333333335, 0.5499999999999999) to[out=0, in=135] (2.5, 0)
to (2.5, 0) to [out=45, in=180] (2.916666666666667, 0.5500000000000002) to[out=0, in=135] (3.3333333333333335, 0)
to (3.3333333333333335, 0) to [out=45, in=180] (3.75, 0.5500000000000002) to[out=0, in=135] (4.166666666666667, 0)
to (4.166666666666667, 0) to [out=45, in=180] (4.583333333333334, 0.5499999999999998) to[out=0, in=135] (5.0, 0)
to (5.0, 0) to [out=45, in=180] (5.416666666666666, 0.5499999999999998) to[out=0, in=135] (5.833333333333333, 0)
to (5.833333333333333, 0) to [out=45, in=180] (6.25, 0.5499999999999998) to[out=0, in=135] (6.666666666666666, 0)
to (6.666666666666666, 0) to [out=45, in=180] (7.083333333333332, 0.5499999999999998) to[out=0, in=135] (7.499999999999999, 0)
to (7.499999999999999, 0) to [out=45, in=180] (7.916666666666666, 0.5499999999999998) to[out=0, in=135] (8.333333333333332, 0)
to (8.333333333333332, 0) to [out=45, in=180] (8.75, 0.5500000000000004) to[out=0, in=135] (9.166666666666666, 0)
to (9.166666666666666, 0) to [out=45, in=180] (9.583333333333332, 0.5500000000000004) to[out=0, in=135] (10.0, 0)
to (10.0, 0) to [out=45, in=180] (10.416666666666668, 0.5500000000000004) to[out=0, in=135] (10.833333333333334, 0)
to (10.833333333333334, 0) to [out=45, in=180] (11.25, 0.5500000000000004) to[out=0, in=135] (11.666666666666668, 0)
to (11.666666666666668, -1.5)
to (0.0, -1.5)
;\draw[densely dotted] (7.083333333333334, -0.5) -- (7.083333333333334, 1.5);
\end{scope}
\end{tikzpicture}

%% file: rys7.tex
\begin{tikzpicture}[scale=0.5]
\useasboundingbox (0,-1) rectangle (10, 4);\draw (0, 0) -- (10, 0);
\draw(1.25, 0.5) node{*};
\begin{scope}
\begin{pgfinterruptboundingbox}
\clip (1.3333333333333335, -1.5) rectangle (7.291666666666667, 1.5);
\end{pgfinterruptboundingbox}
\draw (0.8333333333333334, 0) to[out=45, in=180] (1.25, 0.55) node[above]{} to[out=0, in=135] (1.6666666666666667, 0);
\draw (1.6666666666666667, 0) to[out=45, in=180] (2.0833333333333335, 0.5499999999999999) node[above]{} to[out=0, in=135] (2.5, 0);
\draw (2.5, 0) to[out=45, in=180] (2.916666666666667, 0.5500000000000002) node[above]{} to[out=0, in=135] (3.3333333333333335, 0);
\draw (3.3333333333333335, 0) to[out=45, in=180] (3.75, 0.5500000000000002) node[above]{} to[out=0, in=135] (4.166666666666667, 0);
\draw (4.166666666666667, 0) to[out=45, in=180] (4.583333333333334, 0.5499999999999998) node[above]{} to[out=0, in=135] (5.0, 0);
\draw (5.0, 0) to[out=45, in=180] (5.416666666666666, 0.5499999999999998) node[above]{} to[out=0, in=135] (5.833333333333333, 0);
\draw[thick] (6.25, 0) ellipse (0.4166666666666667 and 0.2);
\draw (5.833333333333333, 0) to[out=45, in=180] (6.25, 0.5499999999999998) node[above]{} to[out=0, in=135] (6.666666666666666, 0);
\draw (6.666666666666666, 0) to[out=45, in=180] (7.083333333333332, 0.5499999999999998) node[above]{} to[out=0, in=135] (7.499999999999999, 0);
\draw (7.499999999999999, 0) to[out=45, in=180] (7.916666666666666, 0.5499999999999998) node[above]{} to[out=0, in=135] (8.333333333333332, 0);
\end{scope}
\draw[dashed] (7.291666666666667, -0.1) -- (7.291666666666667, 0.75) node[above]{\footnotesize $i$};
\begin{scope}
\draw (2.7916666666666665, -1) rectangle ++(4.500000000000001, 0.5) node[pos=.5]{};
\path[clip]
(0.0, -1.5)
to (0.0, 0) to [out=45, in=180] (0.4166666666666667, 0.55) to[out=0, in=135] (0.8333333333333334, 0)
to (0.8333333333333334, 0) to [out=45, in=180] (1.25, 0.55) to[out=0, in=135] (1.6666666666666667, 0)
to (1.6666666666666667, 0) to [out=45, in=180] (2.0833333333333335, 0.5499999999999999) to[out=0, in=135] (2.5, 0)
to (2.5, 0) to [out=45, in=180] (2.916666666666667, 0.5500000000000002) to[out=0, in=135] (3.3333333333333335, 0)
to (3.3333333333333335, 0) to [out=45, in=180] (3.75, 0.5500000000000002) to[out=0, in=135] (4.166666666666667, 0)
to (4.166666666666667, 0) to [out=45, in=180] (4.583333333333334, 0.5499999999999998) to[out=0, in=135] (5.0, 0)
to (5.0, 0) to [out=45, in=180] (5.416666666666666, 0.5499999999999998) to[out=0, in=135] (5.833333333333333, 0)
to (5.833333333333333, 0) to [out=45, in=180] (6.25, 0.5499999999999998) to[out=0, in=135] (6.666666666666666, 0)
to (6.666666666666666, 0) to [out=45, in=180] (7.083333333333332, 0.5499999999999998) to[out=0, in=135] (7.499999999999999, 0)
to (7.499999999999999, 0) to [out=45, in=180] (7.916666666666666, 0.5499999999999998) to[out=0, in=135] (8.333333333333332, 0)
to (8.333333333333332, 0) to [out=45, in=180] (8.75, 0.5500000000000004) to[out=0, in=135] (9.166666666666666, 0)
to (9.166666666666666, 0) to [out=45, in=180] (9.583333333333332, 0.5500000000000004) to[out=0, in=135] (10.0, 0)
to (10.0, 0) to [out=45, in=180] (10.416666666666668, 0.5500000000000004) to[out=0, in=135] (10.833333333333334, 0)
to (10.833333333333334, 0) to [out=45, in=180] (11.25, 0.5500000000000004) to[out=0, in=135] (11.666666666666668, 0)
to (11.666666666666668, -1.5)
to (0.0, -1.5)
;\draw[densely dotted] (2.7916666666666665, -0.5) -- (2.7916666666666665, 1.5);
\end{scope}
\end{tikzpicture}

%% file: rys8.tex
\begin{tikzpicture}[scale=0.5]
\useasboundingbox (0,-1) rectangle (10, 4);\draw (0, 0) -- (10, 0);
\draw(7.166666666666667, 0.5) node{*};
\begin{scope}
\begin{pgfinterruptboundingbox}
\clip (1.958333333333333, -1.5) rectangle (7.083333333333334, 1.5);
\end{pgfinterruptboundingbox}
\draw (1.6666666666666667, 0) to[out=45, in=180] (2.0833333333333335, 0.5499999999999999) node[above]{} to[out=0, in=135] (2.5, 0);
\draw (2.5, 0) to[out=45, in=180] (2.916666666666667, 0.5500000000000002) node[above]{} to[out=0, in=135] (3.3333333333333335, 0);
\draw (3.3333333333333335, 0) to[out=45, in=180] (3.75, 0.5500000000000002) node[above]{} to[out=0, in=135] (4.166666666666667, 0);
\draw (4.166666666666667, 0) to[out=45, in=180] (4.583333333333334, 0.5499999999999998) node[above]{} to[out=0, in=135] (5.0, 0);
\draw[thick] (5.416666666666667, 0) ellipse (0.4166666666666667 and 0.2);
\draw (5.0, 0) to[out=45, in=180] (5.416666666666666, 0.5499999999999998) node[above]{} to[out=0, in=135] (5.833333333333333, 0);
\draw (5.833333333333333, 0) to[out=45, in=180] (6.25, 0.5499999999999998) node[above]{} to[out=0, in=135] (6.666666666666666, 0);
\draw (6.666666666666666, 0) to[out=45, in=180] (7.083333333333332, 0.5499999999999998) node[above]{} to[out=0, in=135] (7.499999999999999, 0);
\draw (7.499999999999999, 0) to[out=45, in=180] (7.916666666666666, 0.5499999999999998) node[above]{} to[out=0, in=135] (8.333333333333332, 0);
\end{scope}
\draw[dashed] (1.958333333333333, -0.1) -- (1.958333333333333, 0.75) node[above]{\footnotesize $j$};
\begin{scope}
\draw (1.958333333333333, -1) rectangle ++(4.500000000000001, 0.5) node[pos=.5]{};
\path[clip]
(0.0, -1.5)
to (0.0, 0) to [out=45, in=180] (0.4166666666666667, 0.55) to[out=0, in=135] (0.8333333333333334, 0)
to (0.8333333333333334, 0) to [out=45, in=180] (1.25, 0.55) to[out=0, in=135] (1.6666666666666667, 0)
to (1.6666666666666667, 0) to [out=45, in=180] (2.0833333333333335, 0.5499999999999999) to[out=0, in=135] (2.5, 0)
to (2.5, 0) to [out=45, in=180] (2.916666666666667, 0.5500000000000002) to[out=0, in=135] (3.3333333333333335, 0)
to (3.3333333333333335, 0) to [out=45, in=180] (3.75, 0.5500000000000002) to[out=0, in=135] (4.166666666666667, 0)
to (4.166666666666667, 0) to [out=45, in=180] (4.583333333333334, 0.5499999999999998) to[out=0, in=135] (5.0, 0)
to (5.0, 0) to [out=45, in=180] (5.416666666666666, 0.5499999999999998) to[out=0, in=135] (5.833333333333333, 0)
to (5.833333333333333, 0) to [out=45, in=180] (6.25, 0.5499999999999998) to[out=0, in=135] (6.666666666666666, 0)
to (6.666666666666666, 0) to [out=45, in=180] (7.083333333333332, 0.5499999999999998) to[out=0, in=135] (7.499999999999999, 0)
to (7.499999999999999, 0) to [out=45, in=180] (7.916666666666666, 0.5499999999999998) to[out=0, in=135] (8.333333333333332, 0)
to (8.333333333333332, 0) to [out=45, in=180] (8.75, 0.5500000000000004) to[out=0, in=135] (9.166666666666666, 0)
to (9.166666666666666, 0) to [out=45, in=180] (9.583333333333332, 0.5500000000000004) to[out=0, in=135] (10.0, 0)
to (10.0, 0) to [out=45, in=180] (10.416666666666668, 0.5500000000000004) to[out=0, in=135] (10.833333333333334, 0)
to (10.833333333333334, 0) to [out=45, in=180] (11.25, 0.5500000000000004) to[out=0, in=135] (11.666666666666668, 0)
to (11.666666666666668, -1.5)
to (0.0, -1.5)
;\draw[densely dotted] (6.458333333333334, -0.5) -- (6.458333333333334, 1.5);
\end{scope}
\end{tikzpicture}

%% file: rys9.tex
\begin{tikzpicture}[scale=0.5]
\useasboundingbox (0,-1) rectangle (10, 4);\draw (0, 0) -- (10, 0);
\begin{scope}
\begin{pgfinterruptboundingbox}
\clip (0.4166666666666667, -1.5) rectangle (6.25, 1.5);
\end{pgfinterruptboundingbox}
\draw (0.0, 0) to[out=45, in=180] (0.4166666666666667, 0.55) node[above]{} to[out=0, in=135] (0.8333333333333334, 0);
\draw (0.8333333333333334, 0) to[out=45, in=180] (1.25, 0.55) node[above]{} to[out=0, in=135] (1.6666666666666667, 0);
\draw (1.6666666666666667, 0) to[out=45, in=180] (2.0833333333333335, 0.5499999999999999) node[above]{} to[out=0, in=135] (2.5, 0);
\draw (2.5, 0) to[out=45, in=180] (2.916666666666667, 0.5500000000000002) node[above]{} to[out=0, in=135] (3.3333333333333335, 0);
\draw (3.3333333333333335, 0) to[out=45, in=180] (3.75, 0.5500000000000002) node[above]{} to[out=0, in=135] (4.166666666666667, 0);
\draw (4.166666666666667, 0) to[out=45, in=180] (4.583333333333334, 0.5499999999999998) node[above]{} to[out=0, in=135] (5.0, 0);
\draw[thick] (5.416666666666667, 0) ellipse (0.4166666666666667 and 0.2);
\draw (5.0, 0) to[out=45, in=180] (5.416666666666666, 0.5499999999999998) node[above]{} to[out=0, in=135] (5.833333333333333, 0);
\draw (5.833333333333333, 0) to[out=45, in=180] (6.25, 0.5499999999999998) node[above]{} to[out=0, in=135] (6.666666666666666, 0);
\draw (6.666666666666666, 0) to[out=45, in=180] (7.083333333333332, 0.5499999999999998) node[above]{} to[out=0, in=135] (7.499999999999999, 0);
\end{scope}
\draw[dashed] (6.25, -0.1) -- (6.25, 0.75) node[above]{\footnotesize $i$};
\begin{scope}
\draw (1.25, -1) rectangle ++(5.0, 0.5) node[pos=.5]{};
\path[clip]
(0.0, -1.5)
to (0.0, 0) to [out=45, in=180] (0.4166666666666667, 0.55) to[out=0, in=135] (0.8333333333333334, 0)
to (0.8333333333333334, 0) to [out=45, in=180] (1.25, 0.55) to[out=0, in=135] (1.6666666666666667, 0)
to (1.6666666666666667, 0) to [out=45, in=180] (2.0833333333333335, 0.5499999999999999) to[out=0, in=135] (2.5, 0)
to (2.5, 0) to [out=45, in=180] (2.916666666666667, 0.5500000000000002) to[out=0, in=135] (3.3333333333333335, 0)
to (3.3333333333333335, 0) to [out=45, in=180] (3.75, 0.5500000000000002) to[out=0, in=135] (4.166666666666667, 0)
to (4.166666666666667, 0) to [out=45, in=180] (4.583333333333334, 0.5499999999999998) to[out=0, in=135] (5.0, 0)
to (5.0, 0) to [out=45, in=180] (5.416666666666666, 0.5499999999999998) to[out=0, in=135] (5.833333333333333, 0)
to (5.833333333333333, 0) to [out=45, in=180] (6.25, 0.5499999999999998) to[out=0, in=135] (6.666666666666666, 0)
to (6.666666666666666, 0) to [out=45, in=180] (7.083333333333332, 0.5499999999999998) to[out=0, in=135] (7.499999999999999, 0)
to (7.499999999999999, 0) to [out=45, in=180] (7.916666666666666, 0.5499999999999998) to[out=0, in=135] (8.333333333333332, 0)
to (8.333333333333332, 0) to [out=45, in=180] (8.75, 0.5500000000000004) to[out=0, in=135] (9.166666666666666, 0)
to (9.166666666666666, 0) to [out=45, in=180] (9.583333333333332, 0.5500000000000004) to[out=0, in=135] (10.0, 0)
to (10.0, 0) to [out=45, in=180] (10.416666666666668, 0.5500000000000004) to[out=0, in=135] (10.833333333333334, 0)
to (10.833333333333334, 0) to [out=45, in=180] (11.25, 0.5500000000000004) to[out=0, in=135] (11.666666666666668, 0)
to (11.666666666666668, -1.5)
to (0.0, -1.5)
;\draw[densely dotted] (1.25, -0.5) -- (1.25, 1.5);
\end{scope}
\end{tikzpicture}

%% file: rys10.tex
\begin{tikzpicture}[scale=0.5]
\useasboundingbox (0,-1) rectangle (10, 4);\draw (0, 0) -- (10, 0);
\begin{scope}
\begin{pgfinterruptboundingbox}
\clip (2.0833333333333335, -1.5) rectangle (7.291666666666667, 1.5);
\end{pgfinterruptboundingbox}
\draw (1.6666666666666667, 0) to[out=45, in=180] (2.0833333333333335, 0.5499999999999999) node[above]{} to[out=0, in=135] (2.5, 0);
\draw (2.5, 0) to[out=45, in=180] (2.916666666666667, 0.5500000000000002) node[above]{} to[out=0, in=135] (3.3333333333333335, 0);
\draw (3.3333333333333335, 0) to[out=45, in=180] (3.75, 0.5500000000000002) node[above]{} to[out=0, in=135] (4.166666666666667, 0);
\draw (4.166666666666667, 0) to[out=45, in=180] (4.583333333333334, 0.5499999999999998) node[above]{} to[out=0, in=135] (5.0, 0);
\draw (5.0, 0) to[out=45, in=180] (5.416666666666666, 0.5499999999999998) node[above]{} to[out=0, in=135] (5.833333333333333, 0);
\draw[thick] (6.25, 0) ellipse (0.4166666666666667 and 0.2);
\draw (5.833333333333333, 0) to[out=45, in=180] (6.25, 0.5499999999999998) node[above]{} to[out=0, in=135] (6.666666666666666, 0);
\draw (6.666666666666666, 0) to[out=45, in=180] (7.083333333333332, 0.5499999999999998) node[above]{} to[out=0, in=135] (7.499999999999999, 0);
\draw (7.499999999999999, 0) to[out=45, in=180] (7.916666666666666, 0.5499999999999998) node[above]{} to[out=0, in=135] (8.333333333333332, 0);
\end{scope}
\draw[dashed] (2.0833333333333335, -0.1) -- (2.0833333333333335, 0.75) node[above]{\footnotesize $j$};
\begin{scope}
\draw (2.0833333333333335, -1) rectangle ++(5.0, 0.5) node[pos=.5]{};
\path[clip]
(0.0, -1.5)
to (0.0, 0) to [out=45, in=180] (0.4166666666666667, 0.55) to[out=0, in=135] (0.8333333333333334, 0)
to (0.8333333333333334, 0) to [out=45, in=180] (1.25, 0.55) to[out=0, in=135] (1.6666666666666667, 0)
to (1.6666666666666667, 0) to [out=45, in=180] (2.0833333333333335, 0.5499999999999999) to[out=0, in=135] (2.5, 0)
to (2.5, 0) to [out=45, in=180] (2.916666666666667, 0.5500000000000002) to[out=0, in=135] (3.3333333333333335, 0)
to (3.3333333333333335, 0) to [out=45, in=180] (3.75, 0.5500000000000002) to[out=0, in=135] (4.166666666666667, 0)
to (4.166666666666667, 0) to [out=45, in=180] (4.583333333333334, 0.5499999999999998) to[out=0, in=135] (5.0, 0)
to (5.0, 0) to [out=45, in=180] (5.416666666666666, 0.5499999999999998) to[out=0, in=135] (5.833333333333333, 0)
to (5.833333333333333, 0) to [out=45, in=180] (6.25, 0.5499999999999998) to[out=0, in=135] (6.666666666666666, 0)
to (6.666666666666666, 0) to [out=45, in=180] (7.083333333333332, 0.5499999999999998) to[out=0, in=135] (7.499999999999999, 0)
to (7.499999999999999, 0) to [out=45, in=180] (7.916666666666666, 0.5499999999999998) to[out=0, in=135] (8.333333333333332, 0)
to (8.333333333333332, 0) to [out=45, in=180] (8.75, 0.5500000000000004) to[out=0, in=135] (9.166666666666666, 0)
to (9.166666666666666, 0) to [out=45, in=180] (9.583333333333332, 0.5500000000000004) to[out=0, in=135] (10.0, 0)
to (10.0, 0) to [out=45, in=180] (10.416666666666668, 0.5500000000000004) to[out=0, in=135] (10.833333333333334, 0)
to (10.833333333333334, 0) to [out=45, in=180] (11.25, 0.5500000000000004) to[out=0, in=135] (11.666666666666668, 0)
to (11.666666666666668, -1.5)
to (0.0, -1.5)
;\draw[densely dotted] (7.083333333333334, -0.5) -- (7.083333333333334, 1.5);
\end{scope}
\end{tikzpicture}